\documentclass{emulateapj}

\usepackage{lscape}
\usepackage{apjfonts}
\usepackage{graphicx}
\usepackage{natbib}
\usepackage{color}
\newcommand{\objs}{SDSS~J0924$+$0510}

\newcommand{\hii}{H{\sevenrm\,II}}
\newcommand{\hbeta}{H{$\beta$}}
\newcommand{\halpha}{H{$\alpha$}}

\newcommand{\hst}{{\it HST}}

\newcommand{\OIII}{[O{\sevenrm\,III}]}

\newcommand{\OIIIb}{[O{\sevenrm\,III}]\,$\lambda$5007}

\newcommand{\NII}{[N{\sevenrm\,II}]}

\newcommand{\SII}{[S{\sevenrm\,II}]}
\newcommand{\SIIa}{[S{\sevenrm\,II}]\,$\lambda$6717}
\newcommand{\SIIb}{[S{\sevenrm\,II}]\,$\lambda$6731}

 \font\sevenrm=cmr7 scaled 1000

\slugcomment{Received 2017 December 4; revised 2018 May 24; accepted 2018 May 30; published 2018 July 19}

\begin{document}

\title{{\it Hubble Space Telescope} Wide Field Camera 3 Identifies an $r_p$=1 Kpc Dual Active Galactic Nucleus in the Minor Galaxy Merger SDSS J0924$+$0510 at $z=0.1495$\altaffilmark{*}}

\shorttitle{\hst\ Identifies A Close Dual AGN}

\shortauthors{LIU ET AL.}
\author{Xin Liu\altaffilmark{1,2}, Hengxiao Guo\altaffilmark{1,2}, Yue Shen\altaffilmark{1,2,4}, Jenny E. Greene\altaffilmark{3}, Michael A. Strauss\altaffilmark{3}}

\altaffiltext{*}{Based, in part, on observations made with the NASA/ESA Hubble Space Telescope, obtained at the Space Telescope Science Institute, which is operated by the Association of Universities for Research in Astronomy, Inc., under NASA contract NAS 5-26555. These observations are associated with program number GO 12521.}

\altaffiltext{1}{Department of Astronomy, University of Illinois at Urbana-Champaign, Urbana, IL 61801, USA; xinliuxl@illinois.edu}

\altaffiltext{2}{National Center for Supercomputing Applications, University of Illinois at Urbana-Champaign, 605 East Springfield Avenue, Champaign, IL 61820, USA}

\altaffiltext{3}{Department of Astrophysical Sciences, Princeton University, Peyton Hall -- Ivy Lane, Princeton, NJ 08544, USA}

\altaffiltext{4}{Alfred P. Sloan Foundation Fellow}

\begin{abstract}
Kiloparsec-scale dual active galactic nuclei (AGNs) are active supermassive black hole (SMBH) pairs co-rotating in galaxies with separations of less than a few kpc.  Expected to be a generic outcome of hierarchical galaxy formation, their frequency and demographics remain uncertain. We have carried out an imaging survey with the Hubble Space Telescope (\hst ) Wide Field Camera 3 (WFC3) of AGNs with double-peaked narrow \OIII\ emission lines. \hst /WFC3 offers high image quality in the near infrared (NIR) to resolve the two stellar nuclei, and in the optical to resolve \OIII\ from ionized gas in the narrow-line regions. This combination has proven to be key in sorting out alternative scenarios. With \hst /WFC3 we are able to explore a new population of close dual AGNs at more advanced merger stages than can be probed from the ground. Here we show that the AGN Sloan Digital Sky Survey (SDSS) J0924$+$0510, which had previously shown two stellar bulges, contains two spatially distinct \OIII\ regions consistent with a dual AGN. While we cannot completely exclude cross-ionization from a single central engine, the nearly equal ratios of \OIII\ strongly suggest a dual AGN with a projected angular separation of 0\farcs4, corresponding to a projected physical separation of $r_p$=1 kpc at redshift $z=0.1495$. This serves as a proof of principle for combining high-resolution NIR and optical imaging to identify close dual AGNs.  Our result suggests that studies based on low-resolution and/or low-sensitivity observations may miss close dual AGNs and thereby may underestimate their occurrence rate on $\lesssim$kpc scales.
\end{abstract}


\section{Introduction}\label{sec:intro}

Galaxy mergers are at the heart of the hierarchical paradigm of galaxy formation \citep{toomre72,Conselice2014}. They transform disks to spheroids, shape the galaxy mass function, and provide routes to star formation \citep{sanders88,hernquist89,perez06,Li2008,jogee09}. Because almost every bulge-dominant galaxy harbors a central supermassive black hole \citep[SMBH;][]{kormendy95,ff05}, mergers will result in the formation of SMBH pairs (separated by $\lesssim$ a few kpc when the black holes (BHs) have not formed a gravitationally bound binary) and gravitationally bound binary SMBHs \citep[usually separated by $\lesssim$ a few pc;][]{begelman80,milosavljevic01,yu02,Khan2013,DEGN}. Because gas-rich mergers are expected to trigger strong gas inflows to the galactic centers \citep{hernquist89,hopkins06}, both BHs in a galaxy merger can simultaneously accrete material and become visible as a dual\footnote{The term ``binary AGN'' was used to describe NGC 6240 \citep[e.g.,][]{komossa03}. There, ``binary'' does not mean that the BHs themselves are gravitationally bound to each other (e.g., $\sim$10 pc for a $\sim$10$^{8}~M_{\odot}$ total BH mass; \citealt{colpi09}). In the case of kpc-scale binary AGNs, the host galaxies dominate the potential well. However, since those early papers, ``binary AGN'' has come to mean those that are gravitationally bound. In this paper we adopt the nomenclature ``dual'' AGNs \citep[e.g.,][]{gerke07,comerford08} or AGN pairs to avoid confusion with binary BHs that are generally considered to be gravitationally bound to each other.} active galactic nucleus (AGN). The demographics of dual AGNs offers a test of the lambda cold dark matter ($\Lambda$CDM) paradigm \citep{volonteri03,yu11,vanwassenhove12,Wang2012,Kulier2015,Steinborn2015,Rosas-Guevara2018,Tremmel2018}.

Dual AGNs also provide a unique probe of the possibly coupled evolution of massive galaxies and their central SMBHs \citep{Foreman2009,colpi09,popovic11,KormendyHo2013}. The tight correlation between the SMBH mass and its host bulge properties \citep{ferrarese00,gebhardt00} suggests that SMBHs may have co-evolved with their host stellar bulges \citep{silk98}. Understanding the physical processes that shape the observed scaling relations between SMBHs and bulges has become a central theme in galaxy formation studies \citep[see][for a recent review on the subject]{KormendyHo2013}. In one of the leading models \citep[e.g.,][]{dimatteo05,hopkins08}, gas-rich mergers trigger strong AGNs that deposit significant energy to the host galaxies and liberate most of the remaining gas, regulating the further growth of both stellar bulges and SMBHs. Simulations suggest that tidal perturbations in the nuclear regions become significant (therefore more likely to trigger nuclear activity and BH inspiral) when the two galaxies (and hence their SMBHs) are separated by $\lesssim$ a few kpc \citep[e.g.,][]{mihos96,hopkins05}. This is also the stage when disks are being turned into bulges, and where the global galactic environment becomes highly perturbed. The rich dynamics and gas physics, and the fact that current facilities can resolve the scales between the two BHs, make dual AGNs a unique laboratory for the study of the effects of merger-induced AGNs on galaxy evolution.

The successive dynamical evolution of SMBH pairs in galaxy mergers is also of great interest \citep{DEGN}. Dual AGNs are the precursors of sub-pc binary SMBHs, with final coalescences that are expected to be a major source of low-frequency gravitational waves \citep{thorne76,vecchio97}. Detailed studies of dual AGNs and their host galaxies \citep[e.g.,][]{max07,Shangguan2016} can address the poorly constrained initial conditions for sub-pc binary SMBHs. This is important for inferring the subsequent accretion and the coupled dynamical evolution \citep{Ivanov1999,Escala2005,Cuadra2009,dotti09,haiman09,Liuyt2010,Kocsis2012a,Mayer2013,Shapiro2013,Farris2015,Shi2015}. A robust evolutionary timescale is needed to enable more realistic forecasts for current and future low-frequency gravitational wave experiments such as pulsar timing arrays \citep[e.g.,][]{jenet04,McWilliams2012,Burke-Spolaor2013,Huerta2015,Shannon2015,Babak2016,Middleton2016,Simon2016,Colpi2017,Mingarelli2017} and space based missions such as the evolved Laser Interferometer Space Antenna \citep[eLISA;][]{haehnelt94,cornish07,trias08,Centrella2010,Klein2016}.  Finally, the frequency of dual AGNs may also provide clues on the physical nature of dark matter particles. For example, in fuzzy dark matter \citep[FDM;][]{Hu2000}, a form of dark matter that consists of extremely light scalar particles with masses on the order of $\sim10^{-22}$ eV, SMBH pairs would never get much closer than $\lesssim1$ kpc because FDM fluctuations may inhibit the orbital decay and inspiral at kiloparsec scales \citep{Hui2017}.

Until a decade ago, only a few unambiguously confirmed dual AGNs with separations of $\lesssim$ a few kpc were known \citep[e.g.,][]{owen85,junkkarinen01,Gregg2002,komossa03,ballo04,bianchi08}. Given their significant importance and apparent scarcity that seems to contradict the naive expectation from $\Lambda$CDM, it is important to build up their statistics and to robustly determine their occurrence rate.  

The past decade has seen significant progress in finding more evidence for dual AGNs \citep[e.g.,][]{comerford09,wang10,comerford11a,Comerford2013,fabbiano11,Koss2011,mazzarella11,Shields2012,tadhunter12,Lena2018}. In particular, systematic searches have achieved orders of magnitude increase in the inventory of dual AGN candidates at (sub-)kpc and tens-of-kpc scales \citep[e.g.,][]{Liu2011a,Fu2018} with a subset of them confirmed in the X-ray \citep[e.g.,][]{green10,koss12,Koss2016,teng12,Liu2013}, radio \citep[e.g.,][]{bondi10,Fu2011a,Deane2014,Muller-Sanchez2015,Fu2015}, or mid-infrared \citep[e.g.,][]{Tsai2013,Ellison2017,Satyapal2017}. Many candidates were selected from AGNs with double-peaked narrow emission lines in the SDSS 3'' fiber aperture \citep[e.g.,][]{wang09,Liu2010b,Smith2010,Ge2012,Barrows2013,Shi2014,LyuLiu2016,Yuan2016}. In these systems, one \OIII\ velocity component is redshifted and the other is blueshifted relative to the systemic velocity (measured from stellar absorption features) by a few hundred km s$^{-1}$. \citet{comerford08} suggested that such systems may be dual AGNs \citep[see also][]{sargent72,heckman81,zhou04,gerke07,xu09,barrows12}, where the two \OIII\ velocity components originate from distinct narrow-line regions (NLRs) around two SMBHs, co-rotating along with their own stellar nuclei in a galaxy merger. Alternatively, these velocity splittings may be caused by NLR kinematics in single AGNs such as galactic disk rotation and/or biconical outflows \citep{Heckman1984,axon98,veilleux01,Greene2005,crenshaw09,rosario10,fischer11,An2013}. Follow-up observations have shown that the majority of AGNs with double-peaked narrow-line profiles are a reflection of NLR kinematics, and $\lesssim$10--20\% are due to dual AGNs \citep{Liu2010a,mcgurk11,Shen2011,Fu2012,Comerford2015,McGurk2015,Villforth2015,Nevin2016}.

\begin{figure*}
  \centering
    \includegraphics[width=170mm]{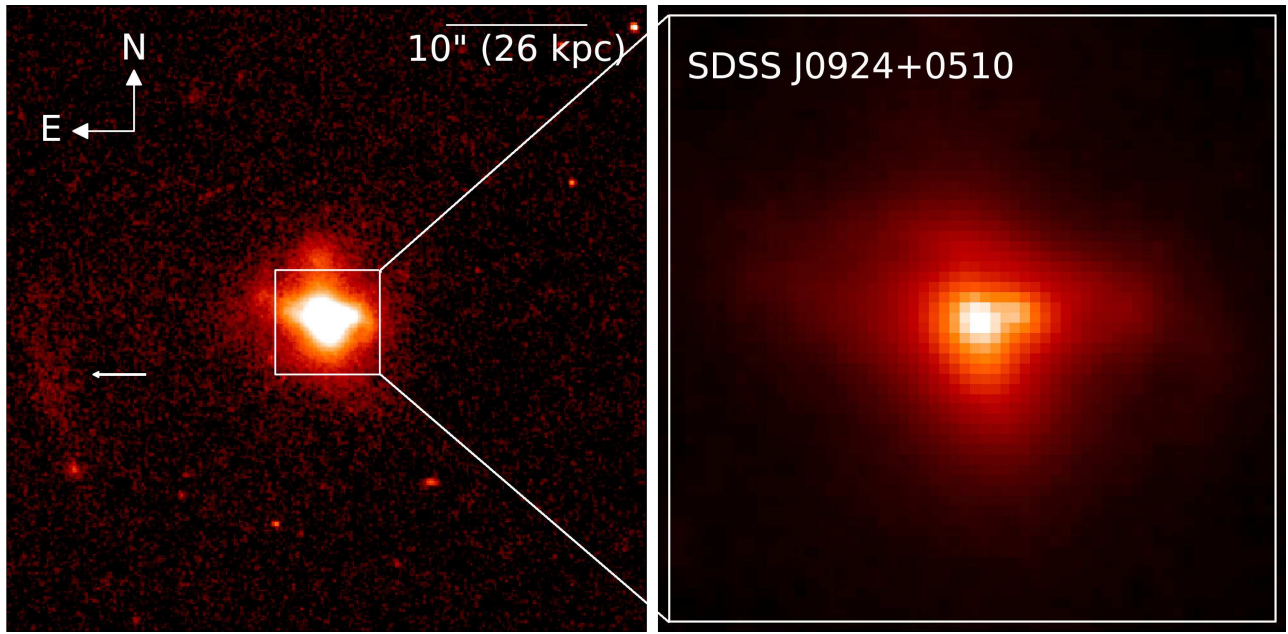}
    \caption{\hst /WFC3 F105W $Y$-band image of our target SDSS J0924$+$0510 at redshift $z=0.1495$. Left: the galaxy merger is in a relatively empty field with no massive companion within at least a 50 kpc radius. In addition to the tidal features seen in the central region of the galaxy merger, there is a low-surface-brightness feature $\sim$50 kpc to the east extending $\sim$30 kpc (indicated with an arrow), which may be a tidal stream related to the merger. The $Y$-band image reaches surface brightness fluctuations of $\mu_{Y} \sim 27.3$ AB mag/arcsec$^{2}$ (1$\sigma$). Right: zoomed-in on the central 8$''\times$8$''$ showing the two stellar nuclei.}
    \label{fig:img}
\end{figure*}

\begin{figure*}
  \centering
    \includegraphics[width=43mm]{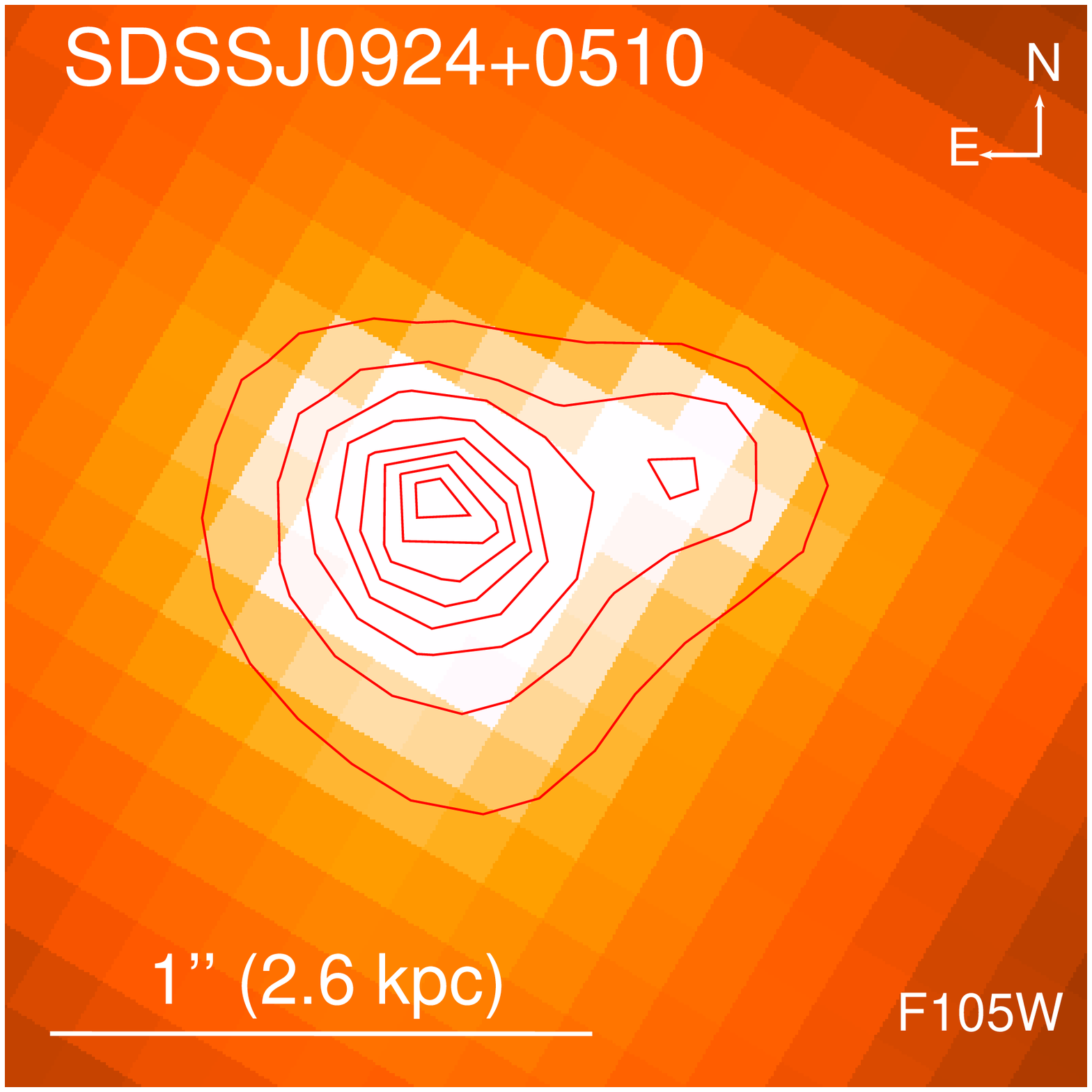}
    \includegraphics[width=43mm]{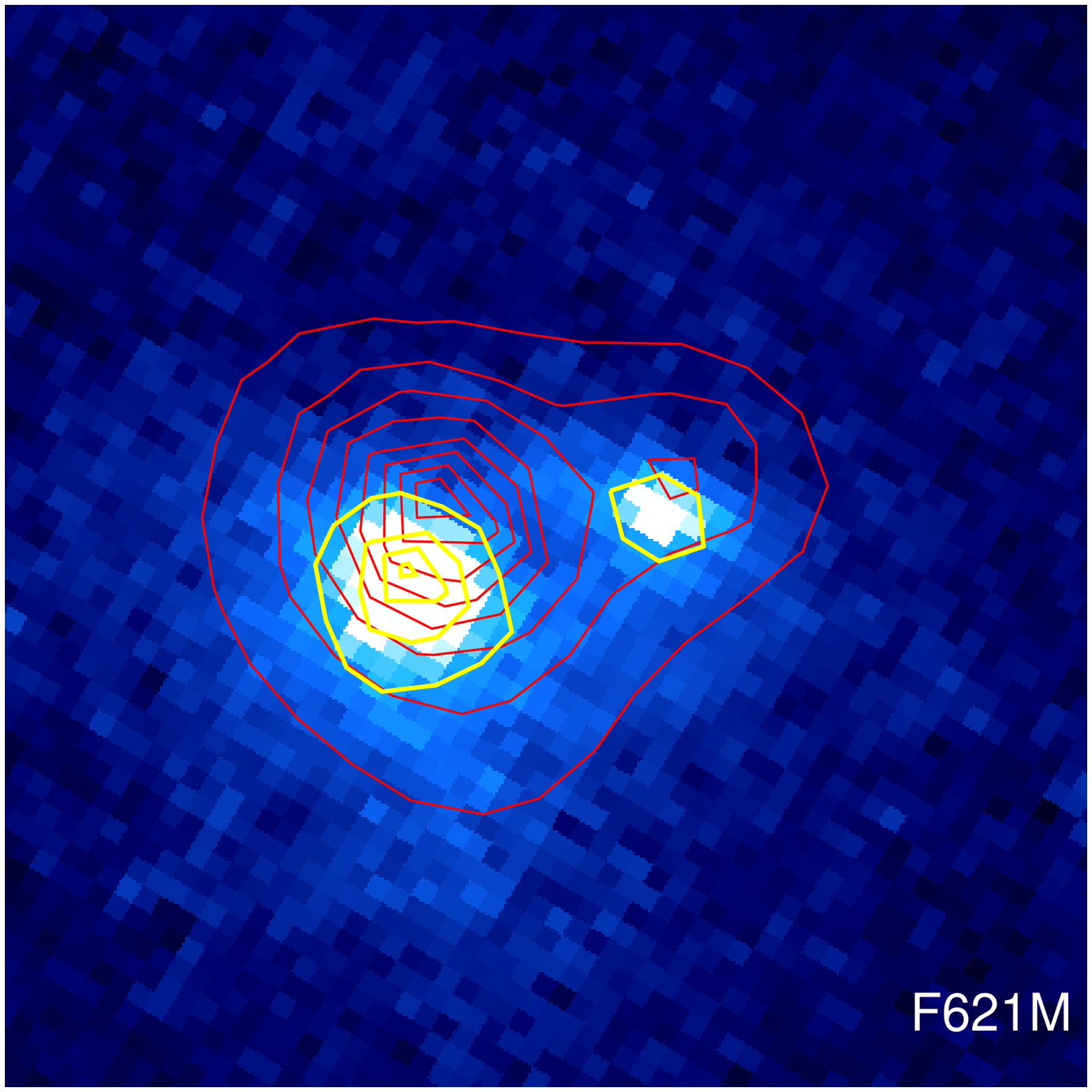}
    \includegraphics[width=43mm]{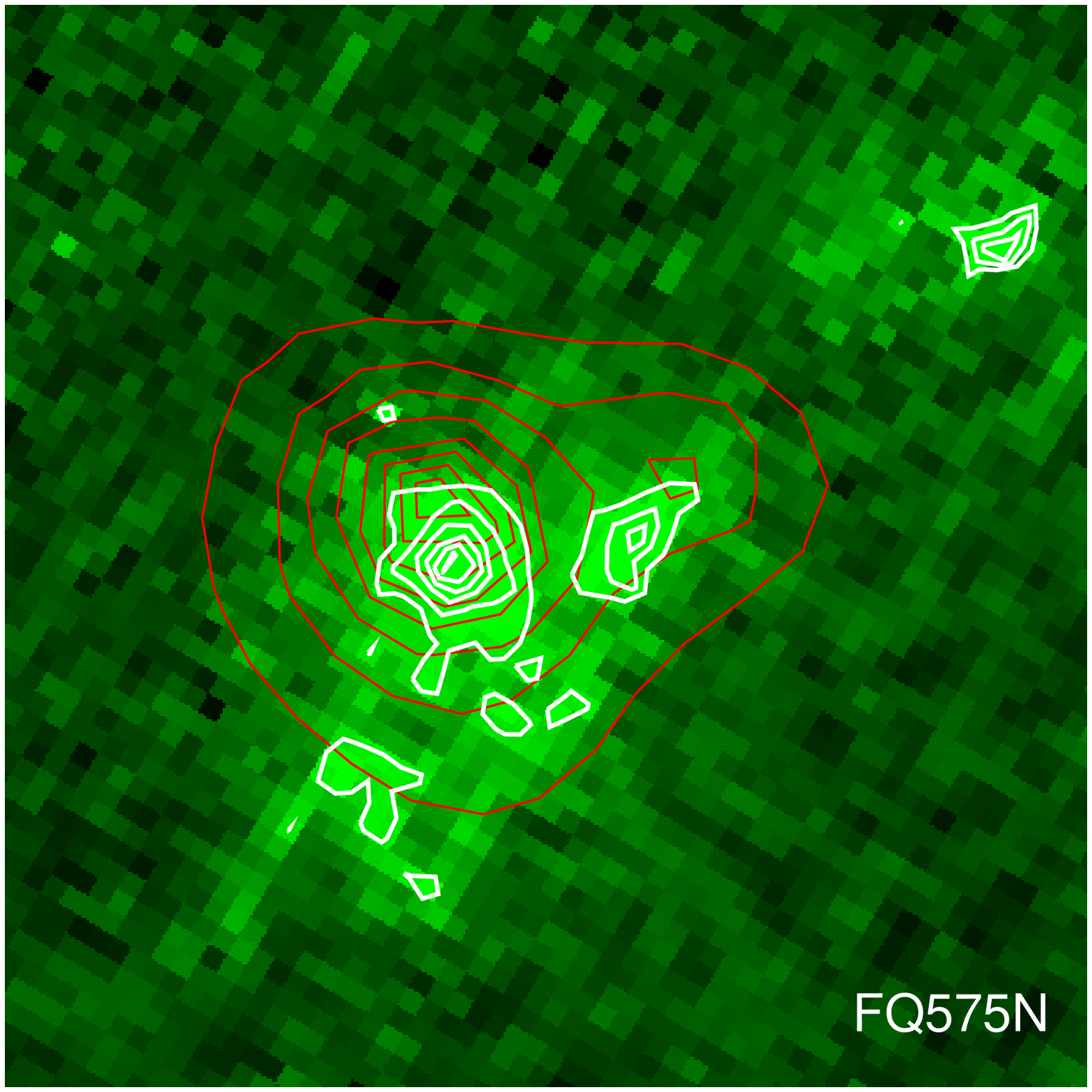}
    \includegraphics[width=43mm]{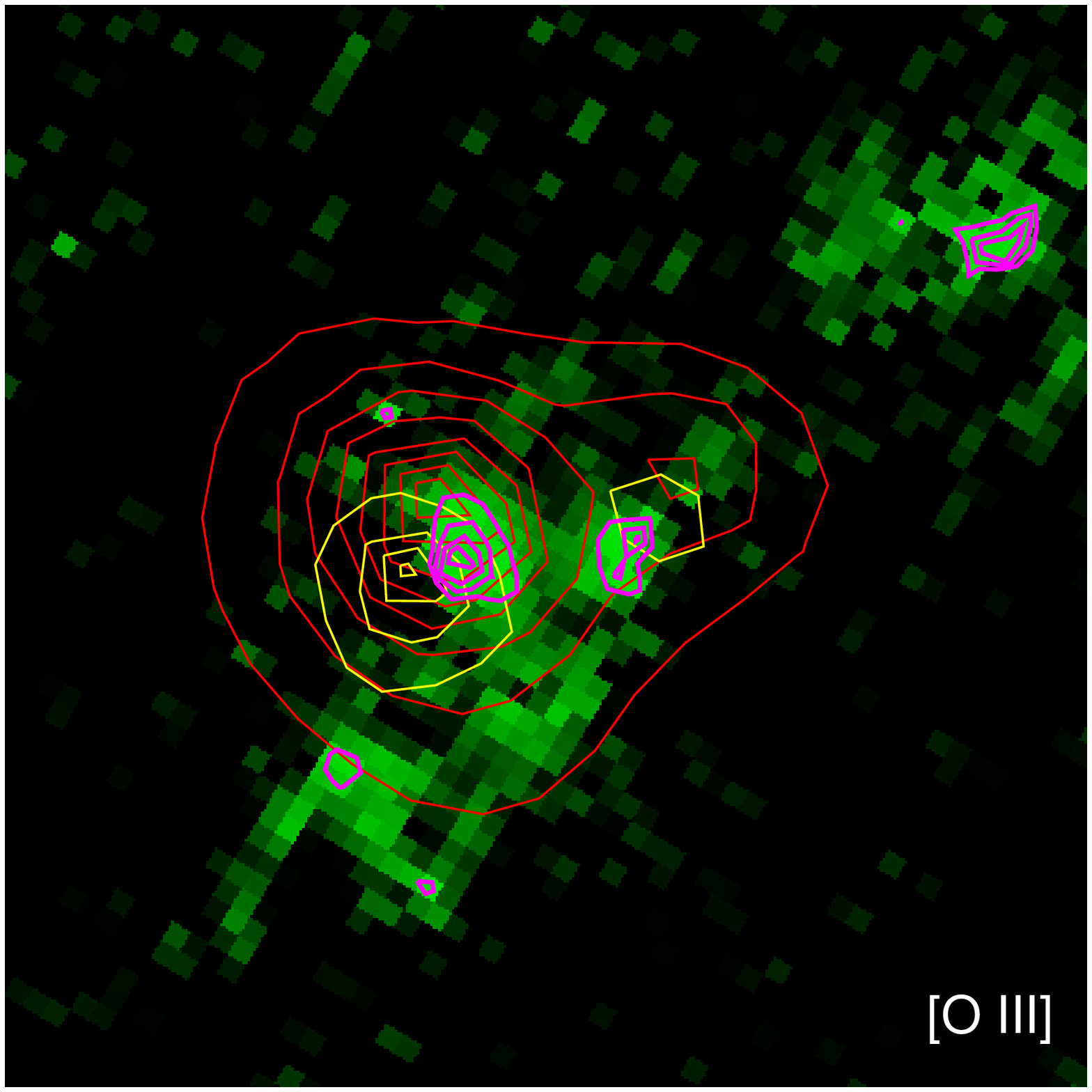}
    \caption{\hst /WFC3 images of SDSS J0924$+$0510 at redshift $z=0.1495$ zoomed in on the central double nucleus. 
    From left to right: F105W ($Y$-band) image showing the two stellar bulges,
    F621M image showing the rest-frame optical stellar continuum just redward of \OIIIb ,
    FQ575N image containing both the \OIIIb\ emission and the stellar continuum, and the continuum-subtracted \OIIIb\ emission image. 
    To guide the eye, we overlay the F105W contours (in red) in all four sub-panels. 
    We show the F621M contours (in yellow) in the F621M and \OIIIb\ sub-panels.
    We also plot contours to highlight the double nucleus seen in the FQ575N (in white) and \OIII -only (in magenta) images. 
    All contours are linearly spaced.
    The double \OIIIb\ nucleus spatially coincides with the F105W double stellar nucleus within the uncertainties in terms of absolute astrometry, 
    although the \OIIIb\ emission is spatially more concentrated than the stars.
    The nucleus separation is smaller (by $\sim$0\farcs1--0\farcs2) in \OIIIb\ emission than in the stellar continuum (see also Figure \ref{fig:spatial}).
    There is also some extended \OIII\ emission to the northwest and southeast of the double nucleus, with no associated overdensity in the stellar continuum images.
    }
    \label{fig:img_zoomin}
\end{figure*}

Using ground-based observations, \citet{Shen2011} have shown that spatially resolved observations for both \OIII\ emission and old stellar bulges are key in discriminating between alternative scenarios \citep[see also][]{comerford11b,Fu2012,McGurk2015,Nevin2016}. In dual AGNs, because the two NLRs are distinct both spatially and in velocity, their dynamics are dominated by the potential of their individual stellar bulges. Hence we expect to see two concentrated \OIII\ nuclei (with extents determined by NLR size convolved with a point-spread function (PSF)) spatially coincident with two stellar bulges in a galaxy merger.  NLR-kinematics systems, on the other hand, will exhibit disk and/or biconical diffuse \OIII\ on top of a smooth, single-peaked stellar background as seen in the prototypical example of Mrk 78 \citep{fischer11}. Despite significant effort in the last few years, the results from many previous studies have been limited by the failure to include both starlight and gas emission and/or the low image quality/sensitivity of the observations.  As we illustrate in this paper, previous studies may have underestimated the frequency of close dual AGNs at $\lesssim$kpc scales and the efficiency of the double-peaked-narrow-line selection technique.

To better determine the frequency of dual AGN and the efficiency of the double-peaked-narrow-line technique in identifying dual AGNs, we have carried out a systematic survey with the Wide Field Camera 3 (WFC3) on board the Hubble Space Telescope (\hst ). To identify dual AGNs, we use UVIS imaging to resolve the double \OIII\ nucleus and $Y$-band imaging to detect the host-galaxy stellar bulges associated with the two \OIII\ nuclei as well as tidal features indicative of galaxy mergers \citep[e.g.,][]{lotz08b}.  {\it HST}/WFC3's high image quality ($\gtrsim$0\farcs1) in both optical and NIR and its high sensitivity allow us to explore dual AGNs at more advanced merger stages. As a proof of concept, here we report the discovery of a dual AGN in the galaxy Sloan Digital Sky Survey (SDSS) J092455.24$+$051052.0 (hereafter \objs\ for short) at redshift $z=0.1495$, with a projected angular separation of 0\farcs4 (corresponding to a projected physical separation of 1 kpc). Figure \ref{fig:img} shows WFC3-IR/$Y$-band imaging of \objs , which reveals a double stellar bulge as well as tidal features indicating an ongoing galaxy merger. Figure \ref{fig:img_zoomin} displays UVIS-F621M and FQ575N images showing two concentrated \OIII\ emission-line nuclei that are spatially coincident with the double stellar bulge, which suggests the presence of two obscured AGNs (see \S \ref{subsec:hst} for details).  While the double stellar bulge has been resolved previously in a ground-based NIR AO image \citep{Fu2012}, the morphology of the \OIII\ emission-line gas was unclear in a previous seeing-limited ground-based integral-field spectrum \citep{Fu2012}. Only with \hst\ are we able to resolve the dual AGN system in \objs\ by combining both the starlight and the emission-line-gas tracers.

We describe our target selection and spectroscopic measurements based on the SDSS spectrum in \S \ref{subsec:target}, the new \hst\ observations and data reduction in \S \ref{subsec:hst}, and data analysis in \S\S \ref{subsec:contsub}--\ref{subsec:o3}, followed by discussion of the nature of the ionizing sources and the origin(s) of the double-peaked \OIII\ emission lines in \objs\ in \S \ref{subsec:result}. We discuss our results and conclude in \S \ref{sec:discuss}.  A Friedmann-Robertson-Walker cosmology with $\Omega_m = 0.3$, $\Omega_{\Lambda} = 0.7$, and $h = 0.7$ is assumed throughout.

\begin{figure}
  \centering
    \includegraphics[width=0.45\textwidth]{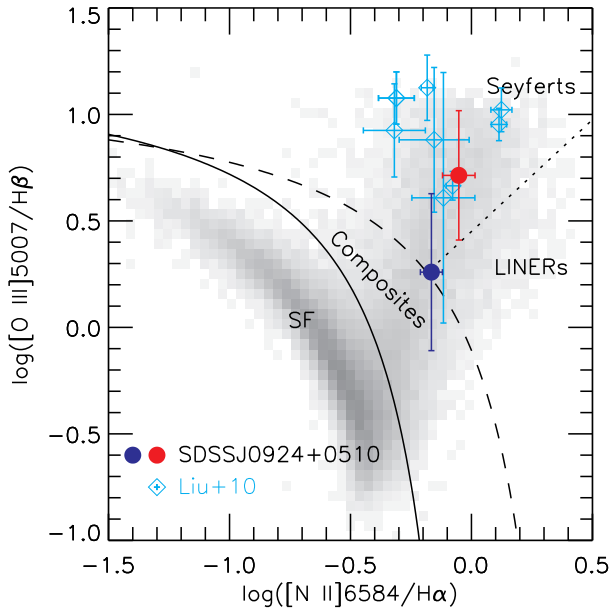}
    \caption{Optical diagnostic emission-line ratios for the target \objs\ measured from the continuum-subtracted SDSS fiber spectrum (Figure \ref{fig:linefit}). The blueshifted (redshifted) velocity component is plotted with a filled circle in blue (red). Also shown for comparison are the four dual AGNs from \citet{Liu2010a}, which were selected from a similar parent sample of Type 2 AGNs with double-peaked narrow emission lines \citep{Liu2010b}, separately for each double-peaked velocity component as measured from spatially resolved longslit spectroscopy. Errorbars denote 1-$\sigma$ statistical errors. The grayscale area represents the number density of 31,179 emission-line galaxies from the SDSS fourth data release \citep[DR4;][]{kauffmann03}. The solid curve is the empirical separation between \hii\ regions and AGNs \citep{kauffmann03}, and the dashed curve is the theoretical ``starburst limit'' \citep{kewley01,Kewley2006}. Pure star-forming (SF) galaxies lie below the solid curve, AGN-dominated galaxies lie above the dashed curve, and AGN-\hii\ composites lie in between. The dotted curve is the empirical separation between Seyferts and Low Ionization Nuclear Emission-line Regions \citep[LINERs;][]{Schawinski2007}.  The redshifted component of \objs\ falls in the Seyfert regime, whereas the blueshifted component sits at the division between Seyferts, LINERs, and AGN-\hii\ composites.}
    \label{fig:bpt}
\end{figure}

\begin{figure}
  \centering
    \includegraphics[width=0.5\textwidth]{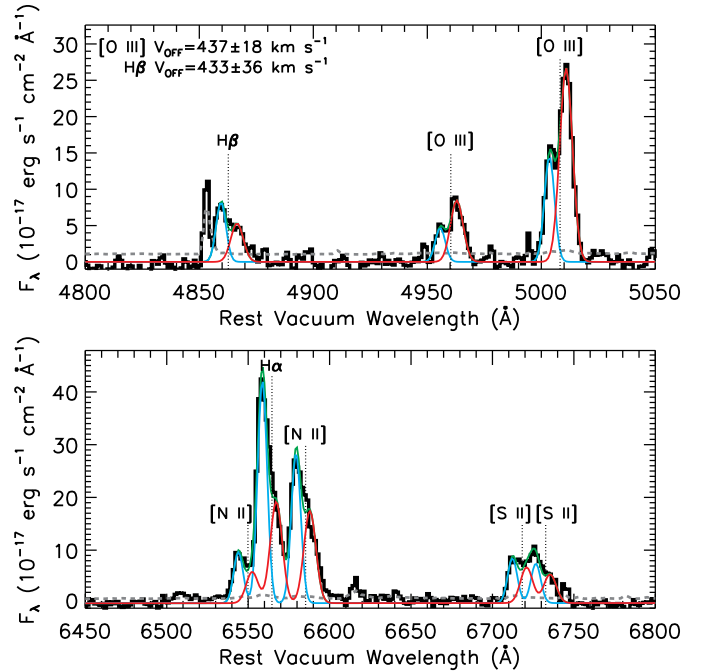}
    \caption{Emission-line models for our target \objs , which exhibits double-peaked narrow emission lines in the continuum-subtracted SDSS fiber spectrum. 
    The upper panel shows the \hbeta --\OIII\ region and the lower panel shows the \halpha --\NII --\SII\ region.
    We plot the data in black and our best-fit model in green. The best-fit model involves two Gaussian components at two different velocities for each emission line, 
    blueshifted (shown in cyan) and redshifted (plotted in red) from the systemic velocity (marked by the dotted vertical lines) as determined from modeling the stellar continuum (Figure \ref{fig:obsspec}). The gray dotted curve denotes the 1-$\sigma$ error in the total flux density spectrum before the continuum subtraction.}
    \label{fig:linefit}
\end{figure}

\section{Observations, Data Reduction, and Data Analysis}\label{sec:obs}

\subsection{Target Selection and Spectroscopic Properties}\label{subsec:target}

Our parent target sample consists of 195 AGNs with double-peaked \OIII\ emission lines with a median redshift $z\sim0.1$. They include both Type 1 and Type 2 sources and were compiled from samples in the literature based on systematic searches \citep{wang09,Liu2010b,Smith2010} using the Seventh Data Release \citep{SDSSDR7} of the SDSS \citep{York2000}. At a completion rate of $\sim$14\%, our \hst\ SNAP program observed 28 out of the 195 approved targets. The SDSS fiber covers 1\farcs5 (3.0 kpc at $z=0.1$) in radius. For the two NLRs to be spatially distinct, they must be separated by $\gtrsim$100 pc due to the intrinsic NLR size given typical \OIII\ luminosity of our targets.  This assumes the luminosity-size relations observed in the \OIII\ emission of AGNs \citep{bennert02,schmitt03,greene11,Hainline2014}. Thus, our sample is well suited for the search of kpc/sub-kpc dual AGNs. Here we focus on the discovery of a close dual AGN candidate in one out of the 28 targets, galaxy \objs , as a proof of concept of our approach. Results on the full observed \hst\ sample will be presented in a future paper. 

The target galaxy \objs\ was first identified by \citet{Smith2010} as a Type 2 AGN that exhibits double-peaked \OIII\ emission lines. It was flagged as a ``Quasi-Stellar Object (QSO)'' by the SDSS spectroscopic pipeline \citep{Bolton2012} because its double-peaked profiles in the narrow emission lines including \hbeta\ mimic ``broad'' emission lines. Figure \ref{fig:bpt} shows that its diagnostic emission-line ratios in both of the double-peaked \OIII\ velocity components either lie in or sit at the boundary of the ``Seyfert'' regime on the Baldwin, Phillips, \& Telervich (BPT) diagram \citep{bpt,veilleux87,kewley01}. This suggests the presence of at least one AGN in the galaxy. We have measured the emission-line ratios based on the SDSS fiber spectrum by decomposing the double-peaked velocity components for the narrow emission lines as shown in Figure \ref{fig:linefit}. We employ a least-$\chi^2$ based parametric model fitting  \citep{Liu2010b}.  As illustrated in Figure \ref{fig:obsspec}, we have carefully subtracted the stellar continuum using a customized code \citep{Liu2009}. We find that the velocity offset between the double-peaked components is $437\pm18$ km s$^{-1}$ in \OIII\ and is $433\pm36$ km s$^{-1}$ in \hbeta\ (Figure \ref{fig:linefit}); i.e., it is consistent with that of \OIII\ within the uncertainties.

\subsection{\hst /WFC3 F105W, F621M, and FQ575N Imaging}\label{subsec:hst}

\begin{figure*}
  \centering
    \includegraphics[width=\textwidth]{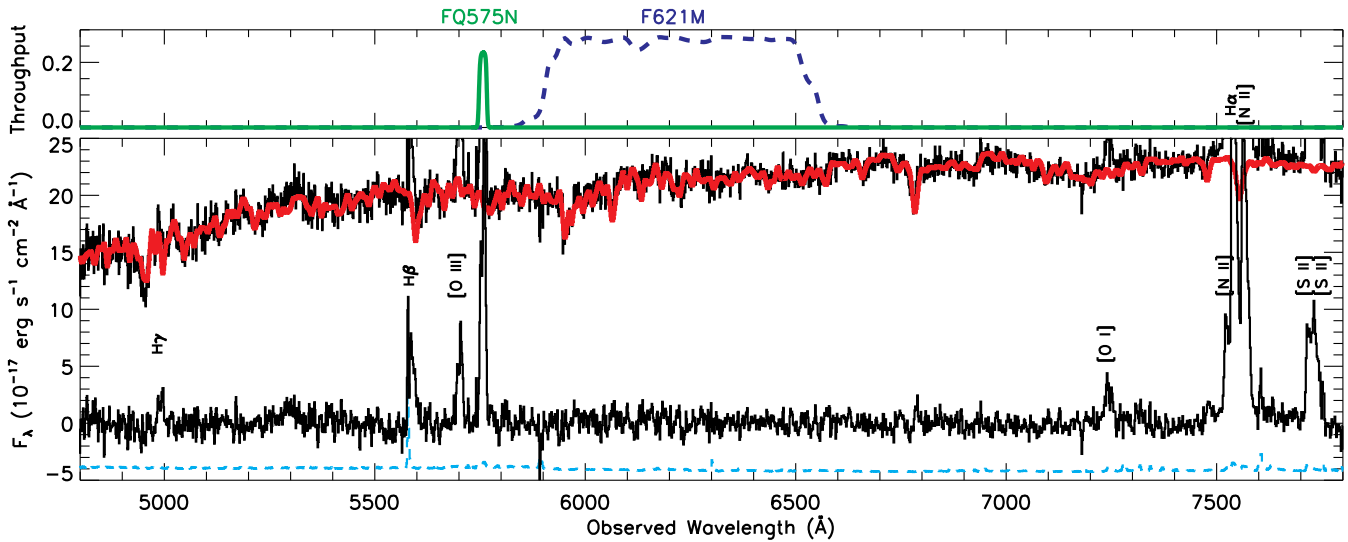}
    \caption{\hst /WFC3-UVIS filter throughputs and stellar continuum modeling for \objs . The lower panel shows the original SDSS fiber spectrum with the best-fit stellar continuum model overlaid on top (plotted in red). Also shown is the emission-line-only residual spectrum with the detected emission lines marked, and the 1-$\sigma$ error spectrum (cyan dashed curve) offset by $-5\times10^{-17}$ erg s$^{-1}$ cm$^{-2}$ \AA$^{-1}$ for clarity. The upper panel displays the system throughput of the two WFC3-UVIS filters that were adopted for the target. The narrow-band filter FQ575N (the green solid curve) covers the \OIIIb\ emission, and the medium-band filter F621M (the blue dashed curve) samples the stellar continuum at rest frame 5920--6520 \AA , just redward of the \OIIIb\ emission line.}
    \label{fig:obsspec}
\end{figure*}

We observed SDSS J0924$+$0510 on 2012 April 18 UT with the WFC3 on board the \hst\ in Cycle 19 (Program: SNAP 12521, PI: X. Liu). It was imaged in the UVIS/F621M (with pivot wavelength\footnote{A measure of the effective wavelength of a filter calculated based on the integrated system throughput \citep{Tokunaga2005}.} $\lambda_{p}=6216.7$ \AA\ and effective width of 631.0 \AA ; \citep{dressel10}), UVIS/FQ575N (with $\lambda_{p}=5755.9$ \AA\ and effective width of 12.9 \AA ), and IR/F105W (wide $Y$-band, with $\lambda_{p}=10489.5$ \AA\ and effective width of 2923.0 \AA ) filters within a single \hst\ orbit. The total net exposure times were 239 s, 100 s, and 900 s in the F105W, F621M, and FQ575N filters, respectively.  The exposure time in the IR-F105W filter was chosen to achieve a signal-to-noise ratio (S/N) of at least 50 per 0\farcs13$\times$0\farcs13 aperture to resolve two stellar nuclei separated by $\gtrsim$0\farcs1 superimposed on a bright galaxy background given the typical brightness of our targets. The exposure times in the UVIS-F621M and UVIS-FQ575N filters were determined by maximizing the S/N in the emission-line detection given a fixed total exposure time as constrained by the length of an \hst\ SNAP orbit after subtracting the exposure time needed by the F105W imaging. 

The F105W filter covers rest frame 7850--10392 \AA , which traces the continuum emission from the old stellar populations in galaxy bulges. Figure \ref{fig:obsspec} shows that the FQ575N filter spans rest frame 5000--5011 \AA , covering the \OIIIb\ line for \objs . An image of the adjacent continua was taken using the F621M filter to measure and subtract the underlying host-galaxy stellar continuum. The WFC3-UVIS charge-coupled device (CCD; IR detector) has a sampling of 0\farcs039 (0\farcs13) pixel$^{-1}$. The F105W (F336W) observations were dithered at four (three) positions to properly sample the IR PSF and to reject cosmic rays and bad pixels. A 1k$\times$1k (512$\times$512) sub-array was employed for the F336W (F105W) imaging, yielding a field of view (FOV) of $40''\times37''$ ($72''\times64''$). 

We reduced the WFC3 data following standard procedures using the \textsf{calwf3} and \textsf{MultiDrizzle} tasks contained in the STSDAS package in PyRAF.  After the \textsf{calwf3} reduction, the images were processed with \textsf{MultiDrizzle} to correct for geometric distortion and pixel area effects. Dithered frames were combined, rejecting cosmic rays and hot pixels. The final image product is calibrated both photometrically and astrometrically. Because \textsf{MultiDrizzle} relies on the measured and catalog positions of guide stars for absolute astrometric calibration, the absolute astrometric accuracy of a WFC3 image processed by \textsf{MultiDrizzle} is limited by the positional uncertainty of guide stars ($\gtrsim$0\farcs2) and the calibration uncertainty of the fine guidance sensor to the instrument aperture ($\sim$0\farcs015). The relative astrometry accuracy of WFC3 images is primarily limited by the uncertainty in the geometric distortion correction of the camera. The typical relative astrometry accuracy is 0\farcs004 for the UVIS and 0\farcs01 for the IR images.

\begin{table*}
\caption{Emission-line Measurements of \objs\ Based on SDSS Spectrum}
\label{table:linefit}
\begin{center}
\scalebox{1.0}{
\begin{tabular}{ccccccccccc}
\hline
\hline
 & V$^{{\rm off}}_{{\rm [O~III]}}$ & V$^{{\rm off}}_{{\rm H}\beta}$ & FWHM$_{{\rm [O~III]}}$ & log F$_{{\rm [O~III]}}$ & log L$_{{\rm [O~III]}}$ & & & log n$_e$ & & $E(B-V)$ \\
Velocity Component & (km s$^{-1}$) & (km s$^{-1}$) & (km s$^{-1}$) & (10$^{-17}$ erg s$^{-1}$ cm$^{-2}$) & (ergs s$^{-1}$) & log(\NII /\halpha) & log(\OIII /\hbeta) & (cm$^{-3}$) & \halpha /\hbeta\ & (mag) \\
(1) & (2) & (3) & (4) & (5) & (6) & (7) & (8) & (9) & (10) & (11) \\ 
\hline
SDSS J0924+0510R & ~~242$\pm$10  & ~~322$\pm$30  & 364$\pm$20 & 2.33 & 41.1 & $-0.05\pm$0.07 & 0.71$\pm$0.30 & 2.34$^{+0.10}_{-0.12}$ & 4.9$\pm$0.6 & 0.54 \\
SDSS J0924+0510B & $-196\pm$15     & $-111\pm$21      & 273$\pm$15 & 1.97 & 40.7 & $-0.17\pm$0.05 & 0.26$\pm$0.37 & 2.55$^{+0.07}_{-0.08}$ & 6.9$\pm$0.8 & 0.89 \\
\hline
\end{tabular}
}
\end{center}
\tablecomments{Columns (2) and (3): velocity offsets relative to the host-galaxy systemic velocity for the redshifted (i.e., \objs\ R) and blueshifted (i.e., \objs\ B) velocity components measured from our best-fit model (Figure \ref{fig:linefit}. Column (4): FWHM of the \OIII\ emission lines measured for each velocity component.  Columns (5) and (6): observed (i.e., uncorrected for intrinsic extinction) \OIIIb\ emission-line flux and luminosity measured from the continuum-subtracted SDSS fiber spectrum; Columns (7) and (8): AGN diagnostic emission-line flux ratios; Column (9): electron density estimated using the diagnostic line ratio \SIIa /\SIIb .
Column (10): emission-line flux ratio;
Column (11): color excess estimated from the emission-line flux ratio \halpha /\hbeta\ (Column 10) using the Balmer decrement method, assuming the intrinsic case B values of 2.87 for $T = 10^4$ K \citep{osterbrock89} and the extinction curve of \citet{cardelli89} with $R_{V} = 3.1$.
}
\end{table*}

\begin{table*}
\caption{\hst /WFC3 F105W and \OIII\ Astrometry and \OIII\ Fluxes of the Double Nucleus in \objs .}
\label{table:result}
\begin{center}
\scalebox{1.0}{
\begin{tabular}{ccccccccc}
\hline
\hline
 & R.A.$_{Y}$ & Decl.$_{Y}$ & R.A.$_{{\rm \OIII }}$ & Decl.$_{{\rm \OIII }}$ & $\Delta\theta_{{\rm diff}}$ & log F$_{{\rm [O~III]}}$ & log L$_{{\rm [O~III]}}$ & R$_{{\rm \OIII }}$   \\
Nucleus Name & (J2000) & (J2000) & (J2000) & (J2000) & ($''$) & (10$^{-17}$ erg s$^{-1}$ cm$^{-2}$) & (erg s$^{-1}$) & (kpc) \\
(1) & (2) & (3) & (4) & (5) & (6) & (7) & (8) & (9) \\ 
\hline
SDSS J0924+0510E  & 09:24:55.279 & +05:10:52.17 & 09:24:55.277 & +05:10:52.09 &  0.09 & 2.19$^{+0.12}_{-0.18}$ & 40.97$^{+0.13}_{-0.17}$ & $<$0.5   \\
SDSS J0924+0510W & 09:24:55.251 & +05:10:52.26 & 09:24:55.255 & +05:10:52.13 &  0.14 & 2.01$^{+0.14}_{-0.20}$ & 40.79$^{+0.14}_{-0.20}$ & $<$0.5 \\
\hline
\end{tabular}}
\end{center}
\tablecomments{Columns (2) and (3): coordinates of the double nucleus measured from \hst\ $Y$-band image. Typical absolute (relative) astrometric uncertainty is $0.''2$ ($0.''01$); Columns (4) and (5): coordinates of the double nucleus measured from continuum-subtracted \OIII\ image; Column (6): difference between the \OIII\ and $Y$-band nucleus positions; Column (7): \OIII\ flux and 1-$\sigma$ statistical uncertainty measured in FQ575N imaging; Column (8): \OIII\ luminosity and 1-$\sigma$ statistical uncertainty measured in FQ575N imaging; Column (9): 3-$\sigma$ upper limit on the \OIII\ size defined as the radius enclosing 50\% of the emission-line flux deconvolved with the PSF.}
\end{table*}

Figures \ref{fig:img}--\ref{fig:img_zoomin} show the F105W, F621M, and FQ575N images of \objs . The zoomed-out view of the F105W image (Figure \ref{fig:img}, left panel) shows tidal features. The zoomed-in view of the F105W image (Figure \ref{fig:img}, right panel) shows two stellar bulges in the central region of the merging galaxy with a projected angular separation of 0\farcs4, corresponding to projected physical separation of just above 1 kpc. The double stellar nucleus and the tidal features suggest that the system is undergoing a merger event.  Because of the small angular separation of the two bulges in this case, only \hst\ or adaptive-optics (AO)-assisted ground-based imaging \citep{Fu2012} can resolve the two components, which would have been missed with lower resolution observations.

\begin{figure*}
\centering
\includegraphics[width=\textwidth]{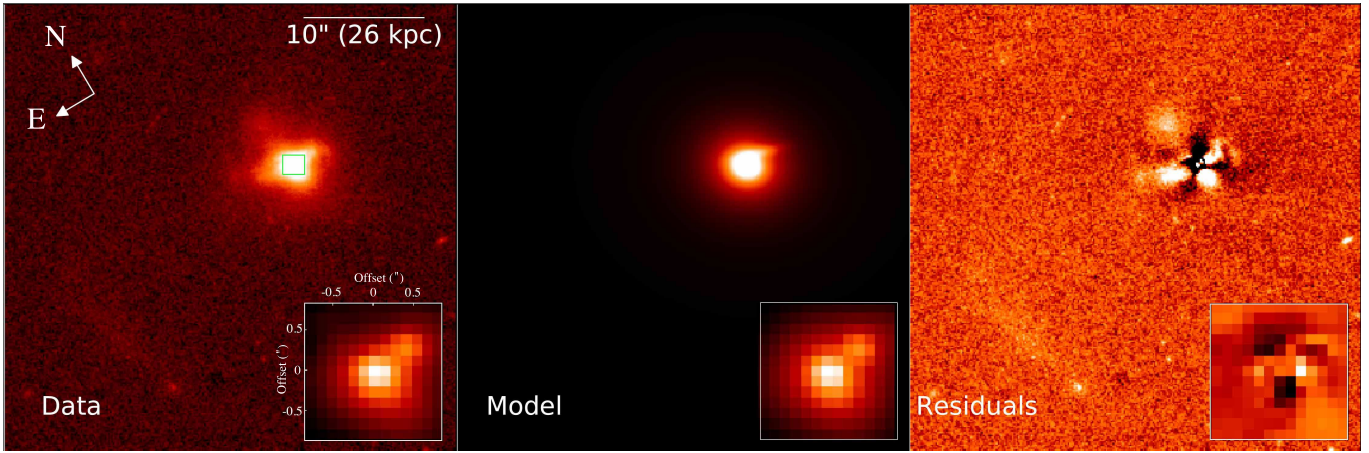}
\caption{Photometric decomposition of the double nucleus in \objs\ from GALFIT analysis of the \hst\ $Y$-band image.
Left: \hst\ $Y$-band image of \objs . 
Middle: our best-fit model from GALFIT analysis consisting of a PSF+S${\rm \acute{e}}$rsic model for each nucleus (Table \ref{table:galfit}). 
Right: residual image, better illustrating tidal features both in the nuclear regions and to the far east of the merger.
We adopt an asinh scale in the data and model images and a linear scale in the residual image.
The inset in each image is zoomed in on the central region with a FOV of 1\farcs6$\times$1\farcs6 centered on the eastern nucleus.
}
\label{fig:dualAGN}
\end{figure*}

\begin{figure*}
\centering
\includegraphics[width=\textwidth]{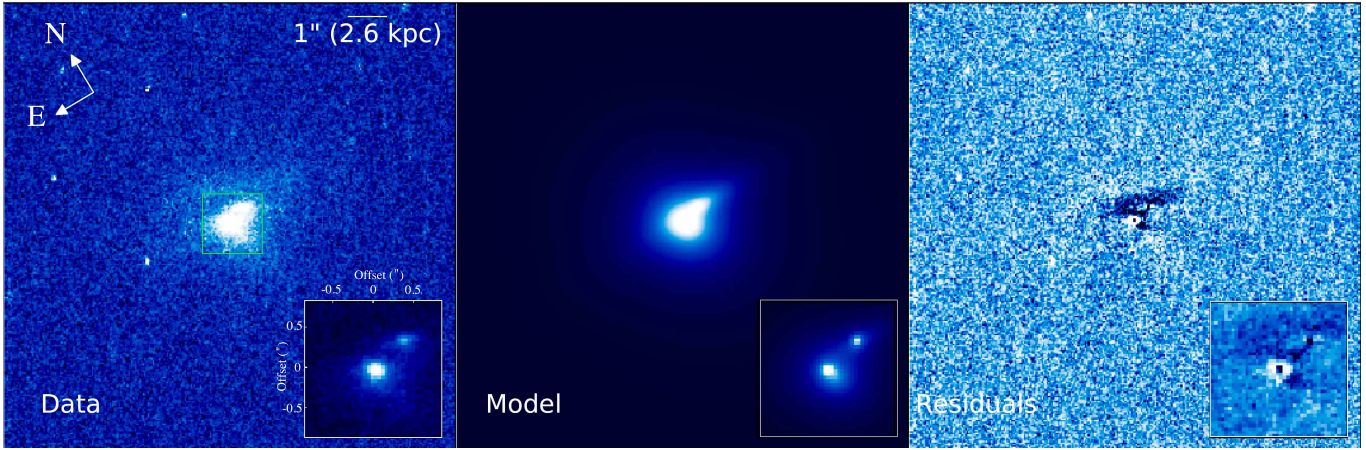}
\caption{Similar to Figure \ref{fig:dualAGN}, but for the F621M-band image decomposition from GALFIT analysis. 
Our best-fit model from GALFIT analysis consists of a PSF+S${\rm \acute{e}}$rsic model for each nucleus (Table \ref{table:galfit}). 
Due to the smaller PSF in the optical, the double nucleus is better resolved in the F621M imaging here than that in the $Y$-band imaging shown in Figure \ref{fig:dualAGN}. 
}
\label{fig:f621m_galfit}
\end{figure*}

\begin{table*}
\caption{Results of Photometric Decomposition of the Double Nucleus in \objs\ from GALFIT Analysis of \hst\ Images}
\begin{center}
\begin{tabular}{cccccccccccc}
  \hline
  \hline
                          & $m_{Y}^{s}$ &$m_{Y}^{p}$&$R_{e, Y}$  &    &   $m_{F621M}^{s}$   &  $m_{F621M}^{p}$  &$R_{e, F621M}$  &    & $M_z$   & $r-z$       & log $M_*$   \\
Nucleus Name  & (mag)            & (mag)  &   (kpc) &  $n_{Y}$ &  (mag)             & (mag)  &   (kpc) &     $n_{F621M}$  & (mag)     &    (mag)        & ($M_{\odot}$)     \\
      (1)               &    (2)              &  (3)              &  (4)            &   (5)  &   (6)  &  (7)      &  (8) & (9) &(10) & (11) &(12)  \\ 
      \hline
SDSS J0924+0510E  & 16.28$\pm$0.01& 19.32$\pm$0.01 & 5.11$\pm0.04$ & 2.85$\pm$0.02 & 17.38$\pm$0.05 & 21.23$\pm$0.05  & 4.13$\pm$0.28 & 2.69$\pm$0.16 & $-$22.65 & 0.75  & 11.2  \\ 
SDSS J0924+0510W & 17.87$\pm$0.02&  20.40$\pm$0.03 & 4.22$\pm$0.05 & 1.50$\pm$0.04  & 19.09$\pm$0.25 & 22.14$\pm$0.09 & 3.1$\pm$1.3 & 1.96$\pm$0.20  & $-$21.04  & 0.87 & 10.6  \\
\hline
\end{tabular}
\end{center}
\tablecomments{Columns (2) and (3): \hst\ $Y$-band apparent magnitude for the S${\rm \acute{e}}$rsic (``s'') and PSF model component (``p''); 
Column (4): $Y$-band effective radius of each nucleus component from the S${\rm \acute{e}}$rsic model;
Column (5): $Y$-band best-fit S${\rm \acute{e}}$rsic index for each nucleus component; 
Columns (6) and (7): \hst\ F621M-band apparent magnitude for the S${\rm \acute{e}}$rsic (``s'') and PSF model component (``p''); 
Column (8): F621M-band effective radius of each nucleus component from the S${\rm \acute{e}}$rsic model;
Column (9): F621M-band best-fit S${\rm \acute{e}}$rsic index for each nucleus component; 
Column (10): SDSS $z$-band absolute magnitude converted from \hst\ $Y$-band magnitude, assuming a flat spectrum; 
Column (11): color index between the SDSS $r$- and $z$-bands estimated using the F621M $-$ $Y$ colors after applying rest-frame k-corrections; 
Column (12): stellar mass for each nucleus calculated using the M/L estimates from the $M_z$ (Column 10) and $r-z$ color (Column 11) assuming the empirical calibration of \citet{bell03}. See \S \ref{subsec:host} for details. The uncertainties listed are reported by GALFIT and should be treated as lower limits to the real uncertainties.
For the eastern nucleus, $M_z$, $r-z$, and $M_*$ are estimated from the PSF+S${\rm \acute{e}}$rsic combined total, whereas they are directly taken from the S${\rm \acute{e}}$rsic model estimates for the western nucleus.}
\label{table:galfit}
\end{table*}

\subsection{Surface Brightness Profile Fitting}

The F105W/$Y$-band image probes old populations in the host-galaxy stellar bulges. It allows us to explore detailed host-galaxy morphology and low-surface-brightness features indicative of mergers. We use GALFIT \citep{peng02,peng10}, a two-dimensional fitting algorithm, to model the multiple structural components in \objs . We aim to decompose the two stellar nuclei and any associated disk components (which may be relevant if disk rotation is the origin of the double-peaked \OIII\ profiles), and to measure low-surface-brightness tidal features in the host galaxy. GALFIT is well suited for these goals. As there are no bright stars within the FOV of our target \objs , we use stars within the FOVs of other targets in our sample to model the PSF. We model the PSF by averaging seven bright but unsaturated stars in the field of our targets\footnote{The \hst\ PSF is known to vary with time so that PSF mismatch may contribute to the systematic uncertainty. To quantify this, we have carried out tests with different PSF models constructed using field stars in different targets in our sample. Our results suggest that the systematic effect due to PSF mismatch is likely to be minor.}.

We adopt the simplest models including a PSF (for any unresolved nucleus) and a single S${\rm \acute{e}}$rsic model (convolved with a PSF) with a variable index $n$ for each merging component. The S${\rm \acute{e}}$rsic model profile is
\begin{equation}
\Sigma(r) = \Sigma_e \, {\rm exp} \bigg [ - \kappa \bigg( \Big ( \frac{r}{r_e} \Big )^{1/n} -1  \bigg ) \bigg ]
\end{equation}
where $\Sigma(r)$ is the pixel surface brightness at radial distance $r$, $\Sigma_e$ is the pixel surface brightness at the effective radius $r_e$, and $\kappa$ is a parameter related to the S${\rm \acute{e}}$rsic index $n$. $n=1$ for an exponential profile, whereas $n=4$ (corresponding to $\kappa=7.67$) for a de Vaucouleurs profile. Bulge-dominated galaxies have high $n$ values (e.g., $n>2$), whereas disk-dominated galaxies have $n$ close to unity.  We tried different combinations of models until we reached the minimum reduced $\chi^2$ for the smallest number of parameters.  We did not fix any of the model parameters except for the model centroid, which is fixed at the peak position of each nucleus. This helps to decompose the two nuclei, in particular for the $Y$-band analysis. A constant sky background was employed in the fits. Our best-fit model contains a PSF+S${\rm \acute{e}}$rsic profile for each nucleus. This is preferred over a simpler model with a S${\rm \acute{e}}$rsic profile for each nucleus (with $\chi^2/\nu = 1.8$ for $\nu=$149 degrees of freedom for the PSF+S${\rm \acute{e}}$rsic model compared to $\chi^2/\nu = 3.4$ for $\nu=$151 degrees of freedom for the S${\rm \acute{e}}$rsic-only model; the $\chi^2/\nu$ values were calculated over the central 1\farcs6$\times$1\farcs6 region as shown in the inset in Figure \ref{fig:dualAGN}), even though the PSF components are much fainter than the S${\rm \acute{e}}$rsic ones (Table \ref{table:galfit}). Physically the PSF components may represent continuum emission from the unresolved NLRs, which are narrow cores in the surface brightness profiles, contributing only a minor fraction to the total light dominated by the host-galaxy starlight. We also tried to include more S${\rm \acute{e}}$rsic model components, but failed to robustly decompose the bulge and disk components for each nucleus due to model degeneracies and the close separation between the two nuclei. 

Figure \ref{fig:dualAGN} shows our best-fit model and the residual of the F105W/$Y$-band image. Tidal features are more clearly seen in the residual map both to the north and the far east of the merging nuclei. We have also carried out surface brightness profile fitting using GALFIT for the F621M image as shown in Figure \ref{fig:f621m_galfit}. This is to estimate the colors and mass-to-light ratios (M/L) of the double nucleus (see \S \ref{subsec:host}). The two nuclei are much better resolved in the F621M band due to its smaller PSF.  Table \ref{table:galfit} lists our best-fit model parameters. Due to degeneracies in the models and systematic effects involved in the data, the listed uncertainties reported by GALFIT should be treated as lower limits to the real uncertainties.

\subsection{Continuum Subtraction}\label{subsec:contsub}

As shown in Figure \ref{fig:obsspec}, the FQ575N filter contains not only \OIIIb\ emission but also stellar continuum emission from the host galaxy. To characterize the spatial distribution of pure \OIIIb\ emission, we first subtract the underlying continuum contained in FQ575N. We estimate the continuum by scaling the F621M image, which samples the adjacent continua. We calculate the scale factor by convolving the SDSS fiber spectrum with the \hst /WFC3 FQ575N and F621M filter response functions. We have neglected any possible spatial variation of the stellar continuum spectral shapes across the FQ575N and F621M filters enclosed within the SDSS 3$''$-diameter fiber coverage. This approximation contributes to the systematic uncertainty of our continuum subtraction. 

Figure \ref{fig:img_zoomin} shows the raw FQ575N image, which contains both \OIIIb emission and stellar continua, the F621M image, and the continuum-subtracted \OIIIb\ image. There is a small offset in the absolute astrometry between the F105W and F621M images so that the peak positions of the two stellar nuclei do not perfectly align across the two images. This is best explained by error in the absolute astrometry. We do not correct for the offset as the two-band positions are consistent with each other within the astrometric uncertainties. We performed aperture photometry using a 3$''$-diameter circle of the continuum-subtracted \OIIIb\ image. The resulting monochromatic \OIIIb\ luminosity density in FQ575N (Table \ref{table:result}) agrees with that inferred from the integrated \OIIIb\ emission-line luminosity measured from the continuum-subtracted SDSS spectrum (Table \ref{table:linefit}) within the uncertainties. This general agreement suggests that the effect of neglecting the spatial variation of the stellar continuum spectral shapes across the FQ575N and F621M filters is small, and that our continuum subtraction was properly done. 

\section{Results}\label{sec:result}

\subsection{Host-galaxy Morphology and Stellar Mass Ratio}\label{subsec:host}

\hst\ $Y$-band images show that the host galaxy is a merger in which the eastern stellar nucleus is $\sim1$ mag brighter than the western stellar nucleus (Figure \ref{fig:dualAGN} and Table \ref{table:galfit}). Our GALFIT analysis suggests that the eastern nucleus is bulge-dominated ($n_Y>2$), whereas the western nucleus is more like disk galaxies ($n_Y<2$). We estimate that the $Y$-band (F621M-band) luminosity ratio between the two stellar nuclei to be $\sim$4.4:1 ($\sim$4.9:1). We estimate stellar masses for the double nucleus based on the $Y$-band magnitudes and F621M $-$ $Y$ colors. We apply k-corrections to convert $m_Y$ and $m_{F621M}$ into SDSS $z$- and $r$-band magnitudes in the rest frame, assuming a flat local spectrum \citep{Blanton2007}. We use $z$-band absolute magnitude and $r-z$ color to estimate stellar masses based on the empirical relation for stellar M/L provided by \citet{bell03},
\begin{equation}
{\rm log}_{10}\left(\frac{M_*}{M_\odot}\right) = - 0.4 \, (M_z - M_{z, \odot}) - 0.041 + 0.463\, (r-z)
\end{equation}
where $M_{*}$ is the galaxy stellar mass in solar units, and $M_{z, \odot}=4.51$ is the absolute magnitude of the Sun in the $z$ band. The estimated stellar mass ratio of the two stellar nuclei is $\sim$4:1, suggesting a minor merger system \citep{Conselice2014}. 

\subsection{Morphology of the {\rm \OIII\ } Emission-line Nuclei and Relation to the Double Stellar Bulge}\label{subsec:o3}

Figure \ref{fig:img_zoomin} shows the morphology of the \OIIIb\ emission-line nuclei. More than $\sim$70\% of the flux is concentrated into two peaks. There is also extended \OIII\ emission both to the northwest and southeast of the two nuclei, although no stellar continuum peak associated with the extended \OIII\ emission is detected.  We have measured the positions of the two \OIIIb\ emission peaks and compared them with those of the stellar continuum peaks (Table \ref{table:result}). The absolute positions of the \OIIIb\ emission peaks are consistent with those of the two stellar nuclei within the astrometric uncertainties. The relative angular separation between the two nuclei, however, is significantly smaller in the \OIIIb\ line emission ($\Delta \theta_{{\rm \OIII }} =$0\farcs28$\pm$0\farcs02) than those in the stellar continua ($\Delta \theta_{{\rm F621M}} =$0\farcs47$\pm$0\farcs02 in F621M and $\Delta \theta_{Y} =$0\farcs45$\pm$0\farcs03 in F105W). This is not surprising considering that some single AGNs do show spatial displacements between the peak in the NLR \OIII\ emission and the host-galaxy stellar bulge center, possibly due to the presence of nuclear star clusters \citep[e.g.,][]{Muller-Sanchez2010,Muller-Sanchez2011}. 

\begin{figure}
  \centering
    \includegraphics[width=0.45\textwidth]{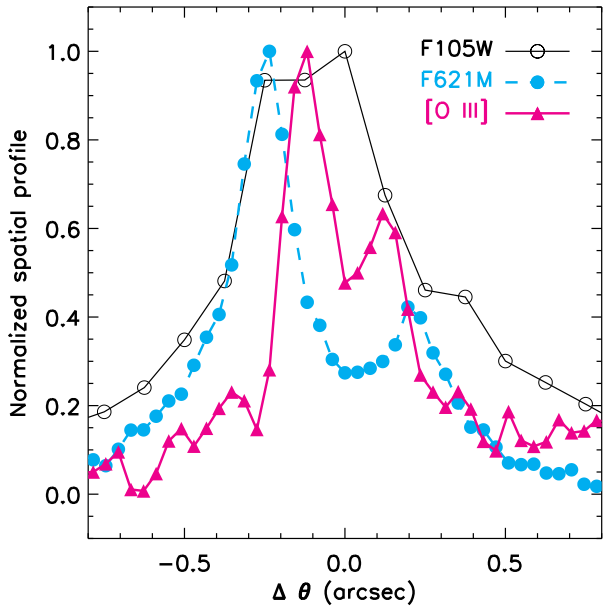}
    \caption{One-dimensional spatial profile projected along PA=96.6 degrees (east of north) of \objs\ in \OIII\ compared with those in the optical (F621M) and $Y$-band (F105W) stellar continua. The spatial profiles were constructed using fluxes extracted from a 0\farcs5$\times$4$''$ box centered at the midpoint between the two nuclei. The systematic offset between the optical (both the F621M continuum shown in cyan filled circles and the \OIII\ emission shown in magenta triangles) and the $Y$-band (shown in black open circles) nucleus peak positions is due to the astrometric errors between the UVIS and IR images, which we do not attempt to correct for (\S \ref{subsec:contsub}). The two \OIII\ nuclei are less separated than those in the stellar continuum and are also spatially more concentrated in each nucleus. See \S \ref{subsec:o3} for details.  }
    \label{fig:spatial}
\end{figure}

The two \OIIIb\ peaks are barely spatially resolved by the \hst\ imaging and are consistent with the PSF. The two \OIII\ nuclei are spatially more concentrated than the two continuum nuclei seen in the $Y$ band (Figure \ref{fig:spatial}). This is expected in the dual AGN scenario, considering that the surface brightness profile of the NLR around each AGN may be spatially more concentrated than that of the stellar continua associated with the stellar bulge. The upper limit on the observed \OIII\ size ($<0.5$ kpc; Table \ref{table:result}) is consistent with the expected typical NLR size expected given the AGN \OIII\ luminosity in each nucleus. Assuming the $R_{{\rm NLR}}$--L$_{{\rm [O\, {\tiny III}]}}$ relation of \citet[][see also \citealt{Bennert2006,greene11,Hainline2013,Hainline2014}]{bennert02}, the expected NLR size is $\sim$0.6 kpc\footnote{Assuming that the $R_{{\rm NLR}}$--L$_{{\rm [O\, {\tiny III}]}}$ relation of \citet{Hainline2013} yields consistent results with our default values after taking into account the difference between $R_{{\rm int}}$ \citep[defined as the size at a limiting surface brightness corrected for cosmological dimming;][]{Liugl2013}, which was adopted by \citet{Hainline2013}, and $R_e$ (i.e., the effective radius enclosing half of the total luminosity) as in our measurement here. $R_{{\rm int}}$ is typically 3--5 times larger than $R_e$ \citep{Liugl2013}. We use $R_e$ because it is relatively insensitive to the detection threshold or to faint extended emission.} ($\sim0.5$ kpc) for the eastern (western) nucleus. The estimated NLR size is indeed smaller than the host-galaxy effective radius that we measure (Table \ref{table:galfit}). 

We estimate the flux ratio between the two \OIIIb\ nuclei to be $\sim1.5\pm0.8$.  This ratio is consistent within the uncertainties with the flux ratio between the two velocity peaks seen in SDSS spectrum ($\sim2.3$; Table \ref{table:linefit}).  Without high-resolution spatially resolved spectroscopy, it is unclear whether there is a one-to-one association between the \OIIIb\ surface brightness peaks seen in imaging and the velocity peaks observed in SDSS spectrum. While the redshifted component is brighter in \OIII\ than the blueshifted component in the spatially integrated spectrum, the \OIII\ fluxes of the two nuclei are consistent within errors, and therefore further high-resolution spatially resolved spectroscopy is still required to confirm which velocity component corresponds to each nucleus. Table \ref{table:result} lists all measurements.

\subsection{Dual AGN, Galactic Disk Rotation, and/or Biconical Outflow?}\label{subsec:result}

We address whether our target is a galaxy merger hosting a dual AGN, a rotating disk, and/or a biconical outflow driven by a single AGN. First, the \hst\ $Y$-band image clearly reveals a double stellar nucleus and tidal features associated with an ongoing merger. The projected angular separation of 0\farcs4 (corresponding to a projected physical separation of 1 kpc at $z=0.1495$) suggests that our target is at a relatively more advanced merger stage than most of the known dual AGNs that have nuclear separations above a few kpc at least. Second, the spatial distribution of the \OIIIb\ emission points to the dual AGN scenario. In the biconical outflow case, one would expect to see diffuse \OIII\ emission against a smooth stellar background. Such systems often show extended \OIII\ emission on both sides of the stellar nucleus, as in Mrk 78 \citep{fischer11}. While a dual AGN and a biconical outflow are not necessarily mutually exclusive \citep[e.g.,][]{Greene2012} and galaxy mergers can trigger both AGN and outflows, it is unlikely that the emission from the outflow gas would be spatially concentrated at the two stellar nuclei in a merger. Finally, the consistent but separately measured velocity offset in the \hbeta\ and \OIII\ emission lines (\S \ref{subsec:target}, Figure \ref{fig:linefit}) suggests a lack of ionization stratification \citep[higher ionization emission lines arise from regions closer to the ionizing sources; e.g., ][]{zamanov02,Komossa2008d} which disfavors a canonical biconical outflow scenario \citep[e.g.,][]{Barrows2013}. 

While single AGNs with rotating disks hosted by isolated non-merging galaxies can also generate double-peaked \OIII\ profiles \citep[e.g.,][]{Shen2011}, our target \objs\ clearly hosts a galaxy merger with two stellar nuclei, both of which are associated with spatially concentrated \OIII\ gas clouds.  However, we cannot rule out the possibility of disk rotation in the gas clouds associated with one or both of the stellar bulges. In the dual AGN scenario, such disk rotation may also contribute substantially to the velocity splitting observed in the spatially integrated SDSS spectrum that caused the selection of the target in the first place. Disk rotation has also been suggested by galaxy merger simulations featuring unusually high disk-to-bulge ratios \citep{Blecha2013a}. 

The case of a dual AGN in \objs , however, is not yet water-tight. Given the proximity of the two nuclei, it is possible that only one BH is actively accreting and ionizing the gas in both components \citep[e.g.,][]{moran92}. High-resolution X-ray and/or radio imaging spectroscopy is needed to further pin down the nature of the ionizing sources in \objs . To resolve the double nucleus and to measure their intrinsic accretion properties, X-ray observations would be extremely challenging even with the superb spatial resolution and sensitivity of Chandra, given the small angular separation between the two nuclei and the likely low count rate in the X-rays as estimated from \OIII\ luminosity \citep{Liu2013}. Our target is not particular radio bright either (covered but undetected in the Faint Images of the Radio Sky at Twenty Centimeters (FIRST) survey; \citealt{white97}), making it more challenging for any future radio confirmation. Future high-resolution, spatially resolved studies of NIR coronal lines (e.g., with integral-field unit spectroscopy with Keck/OSIRIS, VLT/SINFONI, or JWST/NIRSpec) may be able to detect the unambiguous presence of a dual AGN by mapping out the ionization gradients in the merger \citep[e.g.,][]{McGurk2015}.

In conclusion, we suggest that our target \objs\ most likely hosts a dual AGN, although we cannot rule out the possibilities of only one active BH ionizing both galaxies, or the the co-existence of a disk rotation and/or outflow components that may also contribute to the double-peaked \OIII\ profile seen in the spatially integrated SDSS spectrum. Nevertheless, as we have demonstrated here, the combination of high-resolution imaging in the IR and in \OIII\ is a powerful approach to identify close dual AGNs.

\section{Summary and Discussion}\label{sec:discuss}

Kpc-scale dual AGNs are active SMBH pairs co-rotating in merging galaxies with separations of less than a few kpc. Despite decades of searching, and strong theoretical reasons to believe that they are common, the exact frequency and demographics of dual AGNs remain uncertain. We have carried out a snapshot survey with the \hst /WFC3 to systematically identify close dual AGNs. We exploit a well-tested technique based on the selection of AGNs with double-peaked \OIII\ emission lines but have reached the limits of what we can learn from the ground. \hst /WFC3 offers high angular resolution and high-sensitivity imaging both in the NIR (to resolve the old stellar bulges) and in the optical for \OIII\ (to trace the ionized gas in the NLRs), both of which are expected to be associated with the two SMBHs. The combination of observing both the old stellar bulges and the NLR \OIII\ emission has proven to be key in sorting out alternative scenarios \citep[e.g.,][]{Shen2011,Fu2012}. With \hst /WFC3 we are able to explore a new population of close dual AGNs at more advanced merger stages and to better determine the efficiency of the double-peaked selection technique. In this paper we have reported our discovery of a dual AGN with projected angular separation of 0\farcs4 in the galaxy \objs\ at redshift $z=0.1495$, corresponding to a projected physical separation of $r_p$=1 kpc. This serves as a proof of concept for the method of combining high-resolution narrow- and medium-band imaging to identify close dual AGNs.  Although based on only one system, our result has demonstrated that studies based on low-resolution and/or low-sensitivity observations may miss some close dual AGNs, and suggests that they may underestimate the occurrence rate of dual AGNs as well as the efficiency of the double-peaked emission-line selection technique.

The two stellar nuclei in \objs\ were resolved by \citet{Fu2012} using Keck AO-assisted $K_s$-band imaging. The authors categorized \objs\ as ``ambiguous'' since the NLR was unresolved by the seeing-limited optical spectroscopy conducted with the SNIFS on the Hawaiian 2.2 m telescope. As discussed by \citet{Fu2012}, the SNIFS data had lower spatial resolution than that of the AO/\hst\ images, and thus could not resolve structures on scales smaller than $\sim$1\farcs2, so that the authors could not rule out the possibility of emission-line regions that were spatially coincident with the two stellar nuclei. By directly imaging \objs\ both in the NIR and in \OIIIb\ emission in the optical using \hst /WFC3, we have spatially resolved the two NLRs in \objs , thus demonstrating the importance of matching the spatial resolution in \OIII\ imaging for gas with that in the NIR for stars. While it should also be technically feasible, at least in principle, to resolve the weak \OIII\ emission with seeing-limited spatially resolved spectroscopy under good seeing conditions with a larger telescope by employing the technique of spectroastrometry \citep[e.g.,][]{Bailey1998,Whelan2008}, detecting a marginally resolved faint structure on top of a host-galaxy stellar background would be challenging in practice.

It is still an open issue whether the presence of a dual AGN is directly causing the velocity splitting seen in the SDSS spectrum, which caused us to select this target in the first place. Simulations suggest that the velocity field in mergers is likely much more complicated than the simple expectation from our working hypothesis, in which the velocity splitting traces the orbital motion of two NLR clouds co-rotating with two host stellar bulges, in particular in galaxy mergers at more advanced merger phases \citep{Blecha2013a}. High-resolution spatially resolved spectroscopy is needed to determine whether there is a one-to-one association between the \OIIIb\ peaks seen in imaging and the velocity peaks observed in the SDSS spectrum, although such an association is not a requisite condition for the presence of a dual AGN. 

The selection of double-peaked emission line profiles induces a selection bias (e.g., against face-on mergers with smaller line-of-sight (LOS) pairwise velocity offsets). It may not be the most efficient approach in identifying dual AGNs \citep[e.g.,][]{Fu2012}. While the {\it absolute} frequency of dual AGNs may indeed be underestimated using the double-peaked technique if the associated selection bias were not properly accounted for, the {\it relative} frequency of dual AGNs on smaller scales as compared to that on larger scales should not be affected, if the LOS pairwise velocity offset is randomly distributed for galaxy mergers at different phases (on the sub-kpc/kpc scales being considered at least). We can correct for the incompleteness due to the selection of LOS velocity offsets by calibrating against the frequency on larger scales obtained from AGN pair statistics \citep{ellison11,Liu2011a,Liu2012}, which is {\it not} subject to the double-peak selection bias. Taken at face value, the frequency of kpc-scale dual AGNs as inferred from the double-peaked selection sample \citep[e.g., $\sim0.1$\% among all AGNs;][]{Liu2010a,Shen2011,Fu2012} is broadly consistent with that derived based on the AGN pair statistics \citep[$\sim$0.1--0.4\% on kpc scales\footnote{Here it is assumed that the fraction of AGN pairs on kpc scales is related to the fraction on tens-of-kpc scales as $f_{{\rm kpc}} \approx f_{{\rm tens-of-kpc}} \times \tau_{{\rm kpc}}/\tau_{{\rm tens-of-kpc}}$. This in turn assumes that (i) the fraction of observed pairs within a certain range of projected separations scales linearly with the time $\tau$ that a merger spends in that range, and (ii) the probability that two AGNs in a merger are simultaneously active does not strongly depend on the merging phase over the range of separations of a few kpc to tens-of-kpc scales \citep[see Section 4.1 of][]{Liu2011a}.};][]{Liu2011a,Liu2012}.  This general agreement is not particularly surprising, considering that both approaches have selection incompletenesses (e.g., the AGN pair sample of \citealt{Liu2011a} is more incomplete for dual AGNs at $<$5 kpc due to the angular resolution limit of the SDSS imaging). Our optical study is complementary to searches based on AGN selection and dual-AGN identification in the X-rays \citep[e.g.,][]{koss12}, radio \citep[e.g.,][]{Fu2015,Fu2015a}, and mid-IR \citep{Tsai2013,Ellison2017}, although it is not straightforward to directly compare the pairs' statistics across different wavelengths because of various systematic effects that may give rise to some apparent disagreement (e.g., due to differences in sample sizes, AGN populations, merger phases, and host-galaxy or host-halo properties). In future work we aim to address the demographics of our overall \hst\ observed sample in order to directly compare the frequency of dual AGNs as inferred from the double-peaked selection technique with that from the AGN pair statistics. It would also be interesting to carry out more follow-up for new large samples of AGNs with double-peaked narrow emission lines at higher redshift \citep[e.g.,][]{LyuLiu2016,Yuan2016} to address their possible redshift evolution or luminosity dependence \citep[e.g.,][]{yu11,Steinborn2015}.

\acknowledgments

We thank the anonymous referee for a thorough and constructive report that helps improve the paper. Y.S. acknowledges support from the Alfred P. Sloan Foundation and NSF grant 1715579. Support for program number GO 12521 was provided by NASA through a grant from the Space Telescope Science Institute, which is operated by the Association of Universities for Research in Astronomy, Inc., under NASA contract NAS 5-26555.

Funding for the SDSS and SDSS-II has been provided by the Alfred P. Sloan Foundation, the Participating Institutions, the National Science Foundation, the U.S. Department of Energy, the National Aeronautics and Space Administration, the Japanese Monbukagakusho, the Max Planck Society, and the Higher Education Funding Council for England. The SDSS Web Site is http://www.sdss.org/.

Facilities: \hst\ (WFC3), Sloan.

\bibliography{/Users/zeus/Documents/References/binaryrefs}

\begin{thebibliography}{}
\expandafter\ifx\csname natexlab\endcsname\relax\def\natexlab#1{#1}\fi
\providecommand{\url}[1]{\href{#1}{#1}}

\bibitem[{{Abazajian} {et~al.}(2009){Abazajian}, {Adelman-McCarthy},
  {Ag{\"u}eros}, {Allam}, {Allende Prieto}, {An}, {Anderson}, {Anderson},
  {Annis}, {Bahcall}, {Bailer-Jones}, {Barentine}, {Bassett}, {Becker},
  {Beers}, {Bell}, {Belokurov}, {Berlind}, {Berman}, {Bernardi}, {Bickerton},
  {Bizyaev}, {Blakeslee}, {Blanton}, {Bochanski}, {Boroski}, {Brewington},
  {Brinchmann}, {Brinkmann}, {Brunner}, {Budav{\'a}ri}, {Carey}, {Carliles},
  {Carr}, {Castander}, {Cinabro}, {Connolly}, {Csabai}, {Cunha}, {Czarapata},
  {Davenport}, {de Haas}, {Dilday}, {Doi}, {Eisenstein}, {Evans}, {Evans},
  {Fan}, {Friedman}, {Frieman}, {Fukugita}, {G{\"a}nsicke}, {Gates},
  {Gillespie}, {Gilmore}, {Gonzalez}, {Gonzalez}, {Grebel}, {Gunn},
  {Gy{\"o}ry}, {Hall}, {Harding}, {Harris}, {Harvanek}, {Hawley}, {Hayes},
  {Heckman}, {Hendry}, {Hennessy}, {Hindsley}, {Hoblitt}, {Hogan}, {Hogg},
  {Holtzman}, {Hyde}, {Ichikawa}, {Ichikawa}, {Im}, {Ivezi{\'c}}, {Jester},
  {Jiang}, {Johnson}, {Jorgensen}, {Juri{\'c}}, {Kent}, {Kessler}, {Kleinman},
  {Knapp}, {Konishi}, {Kron}, {Krzesinski}, {Kuropatkin}, {Lampeitl},
  {Lebedeva}, {Lee}, {Lee}, {Leger}, {L{\'e}pine}, {Li}, {Lima}, {Lin}, {Long},
  {Loomis}, {Loveday}, {Lupton}, {Magnier}, {Malanushenko}, {Malanushenko},
  {Mandelbaum}, {Margon}, {Marriner}, {Mart{\'{\i}}nez-Delgado}, {Matsubara},
  {McGehee}, {McKay}, {Meiksin}, {Morrison}, {Mullally}, {Munn}, {Murphy},
  {Nash}, {Nebot}, {Neilsen}, {Newberg}, {Newman}, {Nichol}, {Nicinski},
  {Nieto-Santisteban}, {Nitta}, {Okamura}, {Oravetz}, {Ostriker}, {Owen},
  {Padmanabhan}, {Pan}, {Park}, {Pauls}, {Peoples}, {Percival}, {Pier}, {Pope},
  {Pourbaix}, {Price}, {Purger}, {Quinn}, {Raddick}, {Fiorentin}, {Richards},
  {Richmond}, {Riess}, {Rix}, {Rockosi}, {Sako}, {Schlegel}, {Schneider},
  {Scholz}, {Schreiber}, {Schwope}, {Seljak}, {Sesar}, {Sheldon}, {Shimasaku},
  {Sibley}, {Simmons}, {Sivarani}, {Smith}, {Smith}, {Smol{\v c}i{\'c}},
  {Snedden}, {Stebbins}, {Steinmetz}, {Stoughton}, {Strauss}, {Subba Rao},
  {Suto}, {Szalay}, {Szapudi}, {Szkody}, {Tanaka}, {Tegmark}, {Teodoro},
  {Thakar}, {Tremonti}, {Tucker}, {Uomoto}, {Vanden Berk}, {Vandenberg},
  {Vidrih}, {Vogeley}, {Voges}, {Vogt}, {Wadadekar}, {Watters}, {Weinberg},
  {West}, {White}, {Wilhite}, {Wonders}, {Yanny}, {Yocum}, {York}, {Zehavi},
  {Zibetti}, \& {Zucker}}]{SDSSDR7}
{Abazajian}, K.~N., {Adelman-McCarthy}, J.~K., {Ag{\"u}eros}, M.~A., {et~al.}
  2009, \apjs, 182, 543

\bibitem[{{An} {et~al.}(2013){An}, {Paragi}, {Frey}, {Xiao}, {Baan}, {Komossa},
  {Gab{\'a}nyi}, {Xu}, \& {Hong}}]{An2013}
{An}, T., {Paragi}, Z., {Frey}, S., {et~al.} 2013, \mnras, 433, 1161

\bibitem[{{Axon} {et~al.}(1998){Axon}, {Marconi}, {Capetti}, {Maccetto},
  {Schreier}, \& {Robinson}}]{axon98}
{Axon}, D.~J., {Marconi}, A., {Capetti}, A., {et~al.} 1998, \apjl, 496, L75

\bibitem[{{Babak} {et~al.}(2016){Babak}, {Petiteau}, {Sesana}, {Brem},
  {Rosado}, {Taylor}, {Lassus}, {Hessels}, {Bassa}, {Burgay}, {Caballero},
  {Champion}, {Cognard}, {Desvignes}, {Gair}, {Guillemot}, {Janssen},
  {Karuppusamy}, {Kramer}, {Lazarus}, {Lee}, {Lentati}, {Liu}, {Mingarelli},
  {Os{\l}owski}, {Perrodin}, {Possenti}, {Purver}, {Sanidas}, {Smits},
  {Stappers}, {Theureau}, {Tiburzi}, {van Haasteren}, {Vecchio}, \&
  {Verbiest}}]{Babak2016}
{Babak}, S., {Petiteau}, A., {Sesana}, A., {et~al.} 2016, \mnras, 455, 1665

\bibitem[{{Bailey}(1998)}]{Bailey1998}
{Bailey}, J.~A. 1998, in \procspie, Vol. 3355, Optical Astronomical
  Instrumentation, ed. S.~{D'Odorico}, 932--939

\bibitem[{{Baldwin} {et~al.}(1981){Baldwin}, {Phillips}, \& {Terlevich}}]{bpt}
{Baldwin}, J.~A., {Phillips}, M.~M., \& {Terlevich}, R. 1981, \pasp, 93, 5

\bibitem[{{Ballo} {et~al.}(2004){Ballo}, {Braito}, {Della Ceca}, {Maraschi},
  {Tavecchio}, \& {Dadina}}]{ballo04}
{Ballo}, L., {Braito}, V., {Della Ceca}, R., {et~al.} 2004, \apj, 600, 634

\bibitem[{{Barrows} {et~al.}(2013){Barrows}, {Sandberg Lacy}, {Kennefick},
  {Comerford}, {Kennefick}, \& {Berrier}}]{Barrows2013}
{Barrows}, R.~S., {Sandberg Lacy}, C.~H., {Kennefick}, J., {et~al.} 2013, \apj,
  769, 95

\bibitem[{{Barrows} {et~al.}(2012){Barrows}, {Stern}, {Madsen}, {Harrison},
  {Assef}, {Comerford}, {Cushing}, {Fassnacht}, {Gonzalez}, {Griffith},
  {Hickox}, {Kirkpatrick}, \& {Lagattuta}}]{barrows12}
{Barrows}, R.~S., {Stern}, D., {Madsen}, K., {et~al.} 2012, \apj, 744, 7

\bibitem[{{Begelman} {et~al.}(1980){Begelman}, {Blandford}, \&
  {Rees}}]{begelman80}
{Begelman}, M.~C., {Blandford}, R.~D., \& {Rees}, M.~J. 1980, \nat, 287, 307

\bibitem[{{Bell}(2003)}]{bell03}
{Bell}, E.~F. 2003, \apj, 586, 794

\bibitem[{{Bennert} {et~al.}(2002){Bennert}, {Falcke}, {Schulz}, {Wilson}, \&
  {Wills}}]{bennert02}
{Bennert}, N., {Falcke}, H., {Schulz}, H., {Wilson}, A.~S., \& {Wills}, B.~J.
  2002, \apjl, 574, L105

\bibitem[{{Bennert} {et~al.}(2006){Bennert}, {Jungwiert}, {Komossa}, {Haas}, \&
  {Chini}}]{Bennert2006}
{Bennert}, N., {Jungwiert}, B., {Komossa}, S., {Haas}, M., \& {Chini}, R. 2006,
  \aap, 456, 953

\bibitem[{{Bianchi} {et~al.}(2008){Bianchi}, {Chiaberge}, {Piconcelli},
  {Guainazzi}, \& {Matt}}]{bianchi08}
{Bianchi}, S., {Chiaberge}, M., {Piconcelli}, E., {Guainazzi}, M., \& {Matt},
  G. 2008, \mnras, 386, 105

\bibitem[{{Blanton} \& {Roweis}(2007)}]{Blanton2007}
{Blanton}, M.~R., \& {Roweis}, S. 2007, \aj, 133, 734

\bibitem[{{Blecha} {et~al.}(2013){Blecha}, {Loeb}, \& {Narayan}}]{Blecha2013a}
{Blecha}, L., {Loeb}, A., \& {Narayan}, R. 2013, \mnras, 429, 2594

\bibitem[{{Bolton} {et~al.}(2012){Bolton}, {Schlegel}, {Aubourg}, {Bailey},
  {Bhardwaj}, {Brownstein}, {Burles}, {Chen}, {Dawson}, {Eisenstein}, {Gunn},
  {Knapp}, {Loomis}, {Lupton}, {Maraston}, {Muna}, {Myers}, {Olmstead},
  {Padmanabhan}, {P{\^a}ris}, {Percival}, {Petitjean}, {Rockosi}, {Ross},
  {Schneider}, {Shu}, {Strauss}, {Thomas}, {Tremonti}, {Wake}, {Weaver}, \&
  {Wood-Vasey}}]{Bolton2012}
{Bolton}, A.~S., {Schlegel}, D.~J., {Aubourg}, {\'E}., {et~al.} 2012, \aj, 144,
  144

\bibitem[{{Bondi} \& {P{\'e}rez-Torres}(2010)}]{bondi10}
{Bondi}, M., \& {P{\'e}rez-Torres}, M.-A. 2010, \apjl, 714, L271

\bibitem[{{Burke-Spolaor}(2013)}]{Burke-Spolaor2013}
{Burke-Spolaor}, S. 2013, Classical and Quantum Gravity, 30, 224013

\bibitem[{{Cardelli} {et~al.}(1989){Cardelli}, {Clayton}, \&
  {Mathis}}]{cardelli89}
{Cardelli}, J.~A., {Clayton}, G.~C., \& {Mathis}, J.~S. 1989, \apj, 345, 245

\bibitem[{{Centrella} {et~al.}(2010){Centrella}, {Baker}, {Kelly}, \& {van
  Meter}}]{Centrella2010}
{Centrella}, J., {Baker}, J.~G., {Kelly}, B.~J., \& {van Meter}, J.~R. 2010,
  Reviews of Modern Physics, 82, 3069

\bibitem[{{Colpi} \& {Dotti}(2011)}]{colpi09}
{Colpi}, M., \& {Dotti}, M. 2011, Advanced Science Letters, 4, 181

\bibitem[{{Colpi} \& {Sesana}(2017)}]{Colpi2017}
{Colpi}, M., \& {Sesana}, A. 2017, {Gravitational Wave Sources in the Era of
  Multi-Band Gravitational Wave Astronomy}, ed. G.~{Augar} \& E.~{Plagnol}
  (World Scientific Publishing Co), 43--140

\bibitem[{{Comerford} {et~al.}(2012){Comerford}, {Gerke}, {Stern}, {Cooper},
  {Weiner}, {Newman}, {Madsen}, \& {Barrows}}]{comerford11b}
{Comerford}, J.~M., {Gerke}, B.~F., {Stern}, D., {et~al.} 2012, \apj, 753, 42

\bibitem[{{Comerford} {et~al.}(2009{\natexlab{a}}){Comerford}, {Griffith},
  {Gerke}, {Cooper}, {Newman}, {Davis}, \& {Stern}}]{comerford09}
{Comerford}, J.~M., {Griffith}, R.~L., {Gerke}, B.~F., {et~al.}
  2009{\natexlab{a}}, \apjl, 702, L82

\bibitem[{{Comerford} {et~al.}(2015){Comerford}, {Pooley}, {Barrows}, {Greene},
  {Zakamska}, {Madejski}, \& {Cooper}}]{Comerford2015}
{Comerford}, J.~M., {Pooley}, D., {Barrows}, R.~S., {et~al.} 2015, \apj, 806,
  219

\bibitem[{{Comerford} {et~al.}(2011){Comerford}, {Pooley}, {Gerke}, \&
  {Madejski}}]{comerford11a}
{Comerford}, J.~M., {Pooley}, D., {Gerke}, B.~F., \& {Madejski}, G.~M. 2011,
  \apjl, 737, L19

\bibitem[{{Comerford} {et~al.}(2013){Comerford}, {Schluns}, {Greene}, \&
  {Cool}}]{Comerford2013}
{Comerford}, J.~M., {Schluns}, K., {Greene}, J.~E., \& {Cool}, R.~J. 2013,
  \apj, 777, 64

\bibitem[{{Comerford} {et~al.}(2009{\natexlab{b}}){Comerford}, {Gerke},
  {Newman}, {Davis}, {Yan}, {Cooper}, {Faber}, {Koo}, {Coil}, {Rosario}, \&
  {Dutton}}]{comerford08}
{Comerford}, J.~M., {Gerke}, B.~F., {Newman}, J.~A., {et~al.}
  2009{\natexlab{b}}, \apj, 698, 956

\bibitem[{{Conselice}(2014)}]{Conselice2014}
{Conselice}, C.~J. 2014, \araa, 52, 291

\bibitem[{{Cornish} \& {Porter}(2007)}]{cornish07}
{Cornish}, N.~J., \& {Porter}, E.~K. 2007, Classical and Quantum Gravity, 24,
  5729

\bibitem[{{Crenshaw} {et~al.}(2010){Crenshaw}, {Schmitt}, {Kraemer},
  {Mushotzky}, \& {Dunn}}]{crenshaw09}
{Crenshaw}, D.~M., {Schmitt}, H.~R., {Kraemer}, S.~B., {Mushotzky}, R.~F., \&
  {Dunn}, J.~P. 2010, \apj, 708, 419

\bibitem[{{Cuadra} {et~al.}(2009){Cuadra}, {Armitage}, {Alexander}, \&
  {Begelman}}]{Cuadra2009}
{Cuadra}, J., {Armitage}, P.~J., {Alexander}, R.~D., \& {Begelman}, M.~C. 2009,
  \mnras, 393, 1423

\bibitem[{{Deane} {et~al.}(2014){Deane}, {Paragi}, {Jarvis}, {Coriat},
  {Bernardi}, {Fender}, {Frey}, {Heywood}, {Kl{\"o}ckner}, {Grainge}, \&
  {Rumsey}}]{Deane2014}
{Deane}, R.~P., {Paragi}, Z., {Jarvis}, M.~J., {et~al.} 2014, \nat, 511, 57

\bibitem[{{Di Matteo} {et~al.}(2005){Di Matteo}, {Springel}, \&
  {Hernquist}}]{dimatteo05}
{Di Matteo}, T., {Springel}, V., \& {Hernquist}, L. 2005, \nat, 433, 604

\bibitem[{{Dotti} {et~al.}(2009){Dotti}, {Ruszkowski}, {Paredi}, {Colpi},
  {Volonteri}, \& {Haardt}}]{dotti09}
{Dotti}, M., {Ruszkowski}, M., {Paredi}, L., {et~al.} 2009, \mnras, 396, 1640

\bibitem[{{Dressel}(2010)}]{dressel10}
{Dressel}, L. 2010, {Wide Field Camera 3 Instrument Handbook, Version 3.0}, ed.
  {{Dressel}, L.}

\bibitem[{{Ellison} {et~al.}(2011){Ellison}, {Patton}, {Mendel}, \&
  {Scudder}}]{ellison11}
{Ellison}, S.~L., {Patton}, D.~R., {Mendel}, J.~T., \& {Scudder}, J.~M. 2011,
  \mnras, 418, 2043

\bibitem[{{Ellison} {et~al.}(2017){Ellison}, {Secrest}, {Mendel}, {Satyapal},
  \& {Simard}}]{Ellison2017}
{Ellison}, S.~L., {Secrest}, N.~J., {Mendel}, J.~T., {Satyapal}, S., \&
  {Simard}, L. 2017, \mnras, 470, L49

\bibitem[{{Escala} {et~al.}(2005){Escala}, {Larson}, {Coppi}, \&
  {Mardones}}]{Escala2005}
{Escala}, A., {Larson}, R.~B., {Coppi}, P.~S., \& {Mardones}, D. 2005, \apj,
  630, 152

\bibitem[{{Fabbiano} {et~al.}(2011){Fabbiano}, {Wang}, {Elvis}, \&
  {Risaliti}}]{fabbiano11}
{Fabbiano}, G., {Wang}, J., {Elvis}, M., \& {Risaliti}, G. 2011, \nat, 477, 431

\bibitem[{{Farris} {et~al.}(2015){Farris}, {Duffell}, {MacFadyen}, \&
  {Haiman}}]{Farris2015}
{Farris}, B.~D., {Duffell}, P., {MacFadyen}, A.~I., \& {Haiman}, Z. 2015,
  \mnras, 447, L80

\bibitem[{{Ferrarese} \& {Ford}(2005)}]{ff05}
{Ferrarese}, L., \& {Ford}, H. 2005, \ssr, 116, 523

\bibitem[{{Ferrarese} \& {Merritt}(2000)}]{ferrarese00}
{Ferrarese}, L., \& {Merritt}, D. 2000, \apjl, 539, L9

\bibitem[{{Fischer} {et~al.}(2011){Fischer}, {Crenshaw}, {Kraemer}, {Schmitt},
  {Mushotsky}, \& {Dunn}}]{fischer11}
{Fischer}, T.~C., {Crenshaw}, D.~M., {Kraemer}, S.~B., {et~al.} 2011, \apj,
  727, 71

\bibitem[{{Foreman} {et~al.}(2009){Foreman}, {Volonteri}, \&
  {Dotti}}]{Foreman2009}
{Foreman}, G., {Volonteri}, M., \& {Dotti}, M. 2009, \apj, 693, 1554

\bibitem[{{Fu} {et~al.}(2015{\natexlab{a}}){Fu}, {Myers}, {Djorgovski}, {Yan},
  {Wrobel}, \& {Stockton}}]{Fu2015}
{Fu}, H., {Myers}, A.~D., {Djorgovski}, S.~G., {et~al.} 2015{\natexlab{a}},
  \apj, 799, 72

\bibitem[{{Fu} {et~al.}(2015{\natexlab{b}}){Fu}, {Wrobel}, {Myers},
  {Djorgovski}, \& {Yan}}]{Fu2015a}
{Fu}, H., {Wrobel}, J.~M., {Myers}, A.~D., {Djorgovski}, S.~G., \& {Yan}, L.
  2015{\natexlab{b}}, \apjl, 815, L6

\bibitem[{{Fu} {et~al.}(2012){Fu}, {Yan}, {Myers}, {Stockton}, {Djorgovski},
  {Aldering}, \& {Rich}}]{Fu2012}
{Fu}, H., {Yan}, L., {Myers}, A.~D., {et~al.} 2012, \apj, 745, 67

\bibitem[{{Fu} {et~al.}(2011){Fu}, {Zhang}, {Assef}, {Stockton}, {Myers},
  {Yan}, {Djorgovski}, {Wrobel}, \& {Riechers}}]{Fu2011a}
{Fu}, H., {Zhang}, Z.-Y., {Assef}, R.~J., {et~al.} 2011, \apjl, 740, L44

\bibitem[{{Fu} {et~al.}(2018){Fu}, {Steffen}, {Gross}, {Dai}, {Isbell}, {Lin},
  {Wake}, {Xue}, {Bizyaev}, \& {Pan}}]{Fu2018}
{Fu}, H., {Steffen}, J.~L., {Gross}, A.~C., {et~al.} 2018, ArXiv e-prints
  1801.00792, arXiv:1801.00792

\bibitem[{{Ge} {et~al.}(2012){Ge}, {Hu}, {Wang}, {Bai}, \& {Zhang}}]{Ge2012}
{Ge}, J.-Q., {Hu}, C., {Wang}, J.-M., {Bai}, J.-M., \& {Zhang}, S. 2012, \apjs,
  201, 31

\bibitem[{{Gebhardt} {et~al.}(2000){Gebhardt}, {Bender}, {Bower}, {Dressler},
  {Faber}, {Filippenko}, {Green}, {Grillmair}, {Ho}, {Kormendy}, {Lauer},
  {Magorrian}, {Pinkney}, {Richstone}, \& {Tremaine}}]{gebhardt00}
{Gebhardt}, K., {Bender}, R., {Bower}, G., {et~al.} 2000, \apjl, 539, L13

\bibitem[{{Gerke} {et~al.}(2007){Gerke}, {Newman}, {Lotz}, {Yan}, {Barmby},
  {Coil}, {Conselice}, {Ivison}, {Lin}, {Koo}, {Nandra}, {Salim}, {Small},
  {Weiner}, {Cooper}, {Davis}, {Faber}, \& {Guhathakurta}}]{gerke07}
{Gerke}, B.~F., {Newman}, J.~A., {Lotz}, J., {et~al.} 2007, \apjl, 660, L23

\bibitem[{{Green} {et~al.}(2010){Green}, {Myers}, {Barkhouse}, {Mulchaey},
  {Bennert}, {Cox}, \& {Aldcroft}}]{green10}
{Green}, P.~J., {Myers}, A.~D., {Barkhouse}, W.~A., {et~al.} 2010, \apj, 710,
  1578

\bibitem[{{Greene} \& {Ho}(2005)}]{Greene2005}
{Greene}, J.~E., \& {Ho}, L.~C. 2005, \apj, 627, 721

\bibitem[{{Greene} {et~al.}(2011){Greene}, {Zakamska}, {Ho}, \&
  {Barth}}]{greene11}
{Greene}, J.~E., {Zakamska}, N.~L., {Ho}, L.~C., \& {Barth}, A.~J. 2011, \apj,
  732, 9

\bibitem[{{Greene} {et~al.}(2012){Greene}, {Zakamska}, \& {Smith}}]{Greene2012}
{Greene}, J.~E., {Zakamska}, N.~L., \& {Smith}, P.~S. 2012, \apj, 746, 86

\bibitem[{{Gregg} {et~al.}(2002){Gregg}, {Becker}, {White}, {Richards},
  {Chaffee}, \& {Fan}}]{Gregg2002}
{Gregg}, M.~D., {Becker}, R.~H., {White}, R.~L., {et~al.} 2002, \apjl, 573, L85

\bibitem[{{Haehnelt}(1994)}]{haehnelt94}
{Haehnelt}, M.~G. 1994, \mnras, 269, 199

\bibitem[{{Haiman} {et~al.}(2009){Haiman}, {Kocsis}, \& {Menou}}]{haiman09}
{Haiman}, Z., {Kocsis}, B., \& {Menou}, K. 2009, \apj, 700, 1952

\bibitem[{{Hainline} {et~al.}(2013){Hainline}, {Hickox}, {Greene}, {Myers}, \&
  {Zakamska}}]{Hainline2013}
{Hainline}, K.~N., {Hickox}, R., {Greene}, J.~E., {Myers}, A.~D., \&
  {Zakamska}, N.~L. 2013, \apj, 774, 145

\bibitem[{{Hainline} {et~al.}(2014){Hainline}, {Hickox}, {Greene}, {Myers},
  {Zakamska}, {Liu}, \& {Liu}}]{Hainline2014}
{Hainline}, K.~N., {Hickox}, R.~C., {Greene}, J.~E., {et~al.} 2014, \apj, 787,
  65

\bibitem[{{Heckman} {et~al.}(1981){Heckman}, {Miley}, {van Breugel}, \&
  {Butcher}}]{heckman81}
{Heckman}, T.~M., {Miley}, G.~K., {van Breugel}, W.~J.~M., \& {Butcher}, H.~R.
  1981, \apj, 247, 403

\bibitem[{{Heckman} {et~al.}(1984){Heckman}, {van Breugel}, \&
  {Miley}}]{Heckman1984}
{Heckman}, T.~M., {van Breugel}, W.~J.~M., \& {Miley}, G.~K. 1984, \apj, 286,
  509

\bibitem[{{Hernquist}(1989)}]{hernquist89}
{Hernquist}, L. 1989, \nat, 340, 687

\bibitem[{{Hopkins} {et~al.}(2005){Hopkins}, {Hernquist}, {Cox}, {Di Matteo},
  {Robertson}, \& {Springel}}]{hopkins05}
{Hopkins}, P.~F., {Hernquist}, L., {Cox}, T.~J., {et~al.} 2005, \apj, 632, 81

\bibitem[{{Hopkins} {et~al.}(2006){Hopkins}, {Hernquist}, {Cox}, {Di Matteo},
  {Robertson}, \& {Springel}}]{hopkins06}
---. 2006, \apjs, 163, 1

\bibitem[{{Hopkins} {et~al.}(2008){Hopkins}, {Hernquist}, {Cox}, \& {Kere{\v
  s}}}]{hopkins08}
{Hopkins}, P.~F., {Hernquist}, L., {Cox}, T.~J., \& {Kere{\v s}}, D. 2008,
  \apjs, 175, 356

\bibitem[{{Hu} {et~al.}(2000){Hu}, {Barkana}, \& {Gruzinov}}]{Hu2000}
{Hu}, W., {Barkana}, R., \& {Gruzinov}, A. 2000, Physical Review Letters, 85,
  1158

\bibitem[{{Huerta} {et~al.}(2015){Huerta}, {McWilliams}, {Gair}, \&
  {Taylor}}]{Huerta2015}
{Huerta}, E.~A., {McWilliams}, S.~T., {Gair}, J.~R., \& {Taylor}, S.~R. 2015,
  \prd, 92, 063010

\bibitem[{{Hui} {et~al.}(2017){Hui}, {Ostriker}, {Tremaine}, \&
  {Witten}}]{Hui2017}
{Hui}, L., {Ostriker}, J.~P., {Tremaine}, S., \& {Witten}, E. 2017, \prd, 95,
  043541

\bibitem[{{Ivanov} {et~al.}(1999){Ivanov}, {Papaloizou}, \&
  {Polnarev}}]{Ivanov1999}
{Ivanov}, P.~B., {Papaloizou}, J.~C.~B., \& {Polnarev}, A.~G. 1999, \mnras,
  307, 79

\bibitem[{{Jenet} {et~al.}(2004){Jenet}, {Lommen}, {Larson}, \&
  {Wen}}]{jenet04}
{Jenet}, F.~A., {Lommen}, A., {Larson}, S.~L., \& {Wen}, L. 2004, \apj, 606,
  799

\bibitem[{{Jogee} {et~al.}(2009){Jogee}, {Miller}, {Penner}, {Skelton},
  {Conselice}, {Somerville}, {Bell}, {Zheng}, {Rix}, {Robaina}, {Barazza},
  {Barden}, {Borch}, {Beckwith}, {Caldwell}, {Peng}, {Heymans}, {McIntosh},
  {H{\"a}u{\ss}ler}, {Jahnke}, {Meisenheimer}, {Sanchez}, {Wisotzki}, {Wolf},
  \& {Papovich}}]{jogee09}
{Jogee}, S., {Miller}, S.~H., {Penner}, K., {et~al.} 2009, \apj, 697, 1971

\bibitem[{{Junkkarinen} {et~al.}(2001){Junkkarinen}, {Shields}, {Beaver},
  {Burbidge}, {Cohen}, {Hamann}, \& {Lyons}}]{junkkarinen01}
{Junkkarinen}, V., {Shields}, G.~A., {Beaver}, E.~A., {et~al.} 2001, \apjl,
  549, L155

\bibitem[{{Kauffmann} {et~al.}(2003){Kauffmann}, {Heckman}, {Tremonti},
  {Brinchmann}, {Charlot}, {White}, {Ridgway}, {Brinkmann}, {Fukugita}, {Hall},
  {Ivezi{\'c}}, {Richards}, \& {Schneider}}]{kauffmann03}
{Kauffmann}, G., {Heckman}, T.~M., {Tremonti}, C., {et~al.} 2003, \mnras, 346,
  1055

\bibitem[{{Kewley} {et~al.}(2001){Kewley}, {Dopita}, {Sutherland}, {Heisler},
  \& {Trevena}}]{kewley01}
{Kewley}, L.~J., {Dopita}, M.~A., {Sutherland}, R.~S., {Heisler}, C.~A., \&
  {Trevena}, J. 2001, \apj, 556, 121

\bibitem[{{Kewley} {et~al.}(2006){Kewley}, {Groves}, {Kauffmann}, \&
  {Heckman}}]{Kewley2006}
{Kewley}, L.~J., {Groves}, B., {Kauffmann}, G., \& {Heckman}, T. 2006, \mnras,
  372, 961

\bibitem[{{Khan} {et~al.}(2013){Khan}, {Holley-Bockelmann}, {Berczik}, \&
  {Just}}]{Khan2013}
{Khan}, F.~M., {Holley-Bockelmann}, K., {Berczik}, P., \& {Just}, A. 2013,
  \apj, 773, 100

\bibitem[{{Klein} {et~al.}(2016){Klein}, {Barausse}, {Sesana}, {Petiteau},
  {Berti}, {Babak}, {Gair}, {Aoudia}, {Hinder}, {Ohme}, \&
  {Wardell}}]{Klein2016}
{Klein}, A., {Barausse}, E., {Sesana}, A., {et~al.} 2016, \prd, 93, 024003

\bibitem[{{Kocsis} {et~al.}(2012){Kocsis}, {Haiman}, \& {Loeb}}]{Kocsis2012a}
{Kocsis}, B., {Haiman}, Z., \& {Loeb}, A. 2012, \mnras, 427, 2680

\bibitem[{{Komossa} {et~al.}(2003){Komossa}, {Burwitz}, {Hasinger}, {Predehl},
  {Kaastra}, \& {Ikebe}}]{komossa03}
{Komossa}, S., {Burwitz}, V., {Hasinger}, G., {et~al.} 2003, \apjl, 582, L15

\bibitem[{{Komossa} {et~al.}(2008){Komossa}, {Xu}, {Zhou}, {Storchi-Bergmann},
  \& {Binette}}]{Komossa2008d}
{Komossa}, S., {Xu}, D., {Zhou}, H., {Storchi-Bergmann}, T., \& {Binette}, L.
  2008, \apj, 680, 926

\bibitem[{{Kormendy} \& {Ho}(2013)}]{KormendyHo2013}
{Kormendy}, J., \& {Ho}, L.~C. 2013, \araa, 51, 511

\bibitem[{{Kormendy} \& {Richstone}(1995)}]{kormendy95}
{Kormendy}, J., \& {Richstone}, D. 1995, \araa, 33, 581

\bibitem[{{Koss} {et~al.}(2012){Koss}, {Mushotzky}, {Treister}, {Veilleux},
  {Vasudevan}, \& {Trippe}}]{koss12}
{Koss}, M., {Mushotzky}, R., {Treister}, E., {et~al.} 2012, \apjl, 746, L22

\bibitem[{{Koss} {et~al.}(2011){Koss}, {Mushotzky}, {Treister}, {Veilleux},
  {Vasudevan}, {Miller}, {Sanders}, {Schawinski}, \& {Trippe}}]{Koss2011}
---. 2011, \apjl, 735, L42

\bibitem[{{Koss} {et~al.}(2016){Koss}, {Glidden}, {Balokovi{\'c}}, {Stern},
  {Lamperti}, {Assef}, {Bauer}, {Ballantyne}, {Boggs}, {Craig}, {Farrah},
  {F{\"u}rst}, {Gandhi}, {Gehrels}, {Hailey}, {Harrison}, {Markwardt},
  {Masini}, {Ricci}, {Treister}, {Walton}, \& {Zhang}}]{Koss2016}
{Koss}, M.~J., {Glidden}, A., {Balokovi{\'c}}, M., {et~al.} 2016, \apjl, 824,
  L4

\bibitem[{{Kulier} {et~al.}(2015){Kulier}, {Ostriker}, {Natarajan}, {Lackner},
  \& {Cen}}]{Kulier2015}
{Kulier}, A., {Ostriker}, J.~P., {Natarajan}, P., {Lackner}, C.~N., \& {Cen},
  R. 2015, \apj, 799, 178

\bibitem[{{Lena} {et~al.}(2018){Lena}, {Panizo-Espinar}, {Jonker}, {Torres}, \&
  {Heida}}]{Lena2018}
{Lena}, D., {Panizo-Espinar}, G., {Jonker}, P.~G., {Torres}, M., \& {Heida}, M.
  2018, \mnras, arXiv:1804.11232

\bibitem[{{Li} {et~al.}(2008){Li}, {Kauffmann}, {Heckman}, {White}, \&
  {Jing}}]{Li2008}
{Li}, C., {Kauffmann}, G., {Heckman}, T.~M., {White}, S.~D.~M., \& {Jing},
  Y.~P. 2008, \mnras, 385, 1915

\bibitem[{{Liu} {et~al.}(2013{\natexlab{a}}){Liu}, {Zakamska}, {Greene},
  {Nesvadba}, \& {Liu}}]{Liugl2013}
{Liu}, G., {Zakamska}, N.~L., {Greene}, J.~E., {Nesvadba}, N.~P.~H., \& {Liu},
  X. 2013{\natexlab{a}}, \mnras, 430, 2327

\bibitem[{{Liu} {et~al.}(2013{\natexlab{b}}){Liu}, {Civano}, {Shen}, {Green},
  {Greene}, \& {Strauss}}]{Liu2013}
{Liu}, X., {Civano}, F., {Shen}, Y., {et~al.} 2013{\natexlab{b}}, \apj, 762,
  110

\bibitem[{{Liu} {et~al.}(2010{\natexlab{a}}){Liu}, {Greene}, {Shen}, \&
  {Strauss}}]{Liu2010a}
{Liu}, X., {Greene}, J.~E., {Shen}, Y., \& {Strauss}, M.~A. 2010{\natexlab{a}},
  \apjl, 715, L30

\bibitem[{{Liu} {et~al.}(2012){Liu}, {Shen}, \& {Strauss}}]{Liu2012}
{Liu}, X., {Shen}, Y., \& {Strauss}, M.~A. 2012, \apj, 745, 94

\bibitem[{{Liu} {et~al.}(2010{\natexlab{b}}){Liu}, {Shen}, {Strauss}, \&
  {Greene}}]{Liu2010b}
{Liu}, X., {Shen}, Y., {Strauss}, M.~A., \& {Greene}, J.~E. 2010{\natexlab{b}},
  \apj, 708, 427

\bibitem[{{Liu} {et~al.}(2011){Liu}, {Shen}, {Strauss}, \& {Hao}}]{Liu2011a}
{Liu}, X., {Shen}, Y., {Strauss}, M.~A., \& {Hao}, L. 2011, \apj, 737, 101

\bibitem[{{Liu} {et~al.}(2009){Liu}, {Zakamska}, {Greene}, {Strauss}, {Krolik},
  \& {Heckman}}]{Liu2009}
{Liu}, X., {Zakamska}, N.~L., {Greene}, J.~E., {et~al.} 2009, \apj, 702, 1098

\bibitem[{{Liu} \& {Shapiro}(2010)}]{Liuyt2010}
{Liu}, Y.~T., \& {Shapiro}, S.~L. 2010, \prd, 82, 123011

\bibitem[{{Lotz} {et~al.}(2008){Lotz}, {Jonsson}, {Cox}, \&
  {Primack}}]{lotz08b}
{Lotz}, J.~M., {Jonsson}, P., {Cox}, T.~J., \& {Primack}, J.~R. 2008, \mnras,
  391, 1137

\bibitem[{{Lyu} \& {Liu}(2016)}]{LyuLiu2016}
{Lyu}, Y., \& {Liu}, X. 2016, \mnras, 463, 24

\bibitem[{{Max} {et~al.}(2007){Max}, {Canalizo}, \& {de Vries}}]{max07}
{Max}, C.~E., {Canalizo}, G., \& {de Vries}, W.~H. 2007, Science, 316, 1877

\bibitem[{{Mayer}(2013)}]{Mayer2013}
{Mayer}, L. 2013, Classical and Quantum Gravity, 30, 244008

\bibitem[{{Mazzarella} {et~al.}(2012){Mazzarella}, {Iwasawa}, {Vavilkin},
  {Armus}, {Kim}, {Bothun}, {Evans}, {Spoon}, {Haan}, {Howell}, {Lord},
  {Marshall}, {Ishida}, {Xu}, {Petric}, {Sanders}, {Surace}, {Appleton},
  {Chan}, {Frayer}, {Inami}, {Khachikian}, {Madore}, {Privon}, {Sturm}, {U}, \&
  {Veilleux}}]{mazzarella11}
{Mazzarella}, J.~M., {Iwasawa}, K., {Vavilkin}, T., {et~al.} 2012, \aj, 144,
  125

\bibitem[{{McGurk} {et~al.}(2015){McGurk}, {Max}, {Medling}, {Shields}, \&
  {Comerford}}]{McGurk2015}
{McGurk}, R.~C., {Max}, C.~E., {Medling}, A.~M., {Shields}, G.~A., \&
  {Comerford}, J.~M. 2015, \apj, 811, 14

\bibitem[{{McGurk} {et~al.}(2011){McGurk}, {Max}, {Rosario}, {Shields},
  {Smith}, \& {Wright}}]{mcgurk11}
{McGurk}, R.~C., {Max}, C.~E., {Rosario}, D.~J., {et~al.} 2011, \apjl, 738, L2

\bibitem[{{McWilliams} {et~al.}(2012){McWilliams}, {Ostriker}, \&
  {Pretorius}}]{McWilliams2012}
{McWilliams}, S.~T., {Ostriker}, J.~P., \& {Pretorius}, F. 2012, ArXiv e-prints
  1211.4590, arXiv:1211.4590

\bibitem[{Merritt(2013)}]{DEGN}
Merritt, D. 2013, Dynamics and Evolution of Galactic Nuclei, Princeton Series
  in Astrophysics (Princeton University Press).
\newblock \url{http://books.google.com/books?id=cOa1ku640zAC}

\bibitem[{{Middleton} {et~al.}(2016){Middleton}, {Del Pozzo}, {Farr}, {Sesana},
  \& {Vecchio}}]{Middleton2016}
{Middleton}, H., {Del Pozzo}, W., {Farr}, W.~M., {Sesana}, A., \& {Vecchio}, A.
  2016, \mnras, 455, L72

\bibitem[{{Mihos} \& {Hernquist}(1996)}]{mihos96}
{Mihos}, J.~C., \& {Hernquist}, L. 1996, \apj, 464, 641

\bibitem[{{Milosavljevi{\'c}} \& {Merritt}(2001)}]{milosavljevic01}
{Milosavljevi{\'c}}, M., \& {Merritt}, D. 2001, \apj, 563, 34

\bibitem[{{Mingarelli} {et~al.}(2017){Mingarelli}, {Lazio}, {Sesana}, {Greene},
  {Ellis}, {Ma}, {Croft}, {Burke-Spolaor}, \& {Taylor}}]{Mingarelli2017}
{Mingarelli}, C.~M.~F., {Lazio}, T.~J.~W., {Sesana}, A., {et~al.} 2017, ArXiv
  e-prints 1708.03491, arXiv:1708.03491

\bibitem[{{Moran} {et~al.}(1992){Moran}, {Halpern}, {Bothun}, \&
  {Becker}}]{moran92}
{Moran}, E.~C., {Halpern}, J.~P., {Bothun}, G.~D., \& {Becker}, R.~H. 1992,
  \aj, 104, 990

\bibitem[{{M{\"u}ller-S{\'a}nchez} {et~al.}(2015){M{\"u}ller-S{\'a}nchez},
  {Comerford}, {Nevin}, {Barrows}, {Cooper}, \& {Greene}}]{Muller-Sanchez2015}
{M{\"u}ller-S{\'a}nchez}, F., {Comerford}, J.~M., {Nevin}, R., {et~al.} 2015,
  \apj, 813, 103

\bibitem[{{M{\"u}ller-S{\'a}nchez} {et~al.}(2010){M{\"u}ller-S{\'a}nchez},
  {Gonz{\'a}lez-Mart{\'{\i}}n}, {Fern{\'a}ndez-Ontiveros}, {Acosta-Pulido}, \&
  {Prieto}}]{Muller-Sanchez2010}
{M{\"u}ller-S{\'a}nchez}, F., {Gonz{\'a}lez-Mart{\'{\i}}n}, O.,
  {Fern{\'a}ndez-Ontiveros}, J.~A., {Acosta-Pulido}, J.~A., \& {Prieto}, M.~A.
  2010, \apj, 716, 1166

\bibitem[{{M{\"u}ller-S{\'a}nchez} {et~al.}(2011){M{\"u}ller-S{\'a}nchez},
  {Prieto}, {Hicks}, {Vives-Arias}, {Davies}, {Malkan}, {Tacconi}, \&
  {Genzel}}]{Muller-Sanchez2011}
{M{\"u}ller-S{\'a}nchez}, F., {Prieto}, M.~A., {Hicks}, E.~K.~S., {et~al.}
  2011, \apj, 739, 69

\bibitem[{{Nevin} {et~al.}(2016){Nevin}, {Comerford}, {M{\"u}ller-S{\'a}nchez},
  {Barrows}, \& {Cooper}}]{Nevin2016}
{Nevin}, R., {Comerford}, J., {M{\"u}ller-S{\'a}nchez}, F., {Barrows}, R., \&
  {Cooper}, M. 2016, \apj, 832, 67

\bibitem[{{Osterbrock}(1989)}]{osterbrock89}
{Osterbrock}, D.~E. 1989, {Astrophysics of gaseous nebulae and active galactic
  nuclei} (Research supported by the University of California, John Simon
  Guggenheim Memorial Foundation, University of Minnesota, et al.~Mill Valley,
  CA, University Science Books, 1989, 422 p.)

\bibitem[{{Owen} {et~al.}(1985){Owen}, {O'Dea}, {Inoue}, \& {Eilek}}]{owen85}
{Owen}, F.~N., {O'Dea}, C.~P., {Inoue}, M., \& {Eilek}, J.~A. 1985, \apjl, 294,
  L85

\bibitem[{{Peng} {et~al.}(2002){Peng}, {Ho}, {Impey}, \& {Rix}}]{peng02}
{Peng}, C.~Y., {Ho}, L.~C., {Impey}, C.~D., \& {Rix}, H.-W. 2002, \aj, 124, 266

\bibitem[{{Peng} {et~al.}(2010){Peng}, {Ho}, {Impey}, \& {Rix}}]{peng10}
---. 2010, \aj, 139, 2097

\bibitem[{{Perez} {et~al.}(2006){Perez}, {Tissera}, {Lambas}, \&
  {Scannapieco}}]{perez06}
{Perez}, M.~J., {Tissera}, P.~B., {Lambas}, D.~G., \& {Scannapieco}, C. 2006,
  \aap, 449, 23

\bibitem[{{Popovi{\'c}}(2012)}]{popovic11}
{Popovi{\'c}}, L.~{\v C}. 2012, NewAR, 56, 74

\bibitem[{{Rosario} {et~al.}(2010){Rosario}, {Shields}, {Taylor}, {Salviander},
  \& {Smith}}]{rosario10}
{Rosario}, D.~J., {Shields}, G.~A., {Taylor}, G.~B., {Salviander}, S., \&
  {Smith}, K.~L. 2010, \apj, 716, 131

\bibitem[{{Rosas-Guevara} {et~al.}(2018){Rosas-Guevara}, {Bower}, {McAlpine},
  {Bonoli}, \& {Tissera}}]{Rosas-Guevara2018}
{Rosas-Guevara}, Y., {Bower}, R., {McAlpine}, S., {Bonoli}, S., \& {Tissera},
  P. 2018, ArXiv e-prints 1805.01479, arXiv:1805.01479

\bibitem[{{Sanders} {et~al.}(1988){Sanders}, {Soifer}, {Elias}, {Madore},
  {Matthews}, {Neugebauer}, \& {Scoville}}]{sanders88}
{Sanders}, D.~B., {Soifer}, B.~T., {Elias}, J.~H., {et~al.} 1988, \apj, 325, 74

\bibitem[{{Sargent}(1972)}]{sargent72}
{Sargent}, W.~L.~W. 1972, \apj, 173, 7

\bibitem[{{Satyapal} {et~al.}(2017){Satyapal}, {Secrest}, {Ricci}, {Ellison},
  {Rothberg}, {Blecha}, {Constantin}, {Gliozzi}, {McNulty}, \&
  {Ferguson}}]{Satyapal2017}
{Satyapal}, S., {Secrest}, N.~J., {Ricci}, C., {et~al.} 2017, \apj, 848, 126

\bibitem[{{Schawinski} {et~al.}(2007){Schawinski}, {Thomas}, {Sarzi},
  {Maraston}, {Kaviraj}, {Joo}, {Yi}, \& {Silk}}]{Schawinski2007}
{Schawinski}, K., {Thomas}, D., {Sarzi}, M., {et~al.} 2007, \mnras, 382, 1415

\bibitem[{{Schmitt} {et~al.}(2003){Schmitt}, {Donley}, {Antonucci},
  {Hutchings}, {Kinney}, \& {Pringle}}]{schmitt03}
{Schmitt}, H.~R., {Donley}, J.~L., {Antonucci}, R.~R.~J., {et~al.} 2003, \apj,
  597, 768

\bibitem[{{Shangguan} {et~al.}(2016){Shangguan}, {Liu}, {Ho}, {Shen}, {Peng},
  {Greene}, \& {Strauss}}]{Shangguan2016}
{Shangguan}, J., {Liu}, X., {Ho}, L.~C., {et~al.} 2016, \apj, 823, 50

\bibitem[{{Shannon} {et~al.}(2015){Shannon}, {Ravi}, {Lentati}, {Lasky},
  {Hobbs}, {Kerr}, {Manchester}, {Coles}, {Levin}, {Bailes}, {Bhat},
  {Burke-Spolaor}, {Dai}, {Keith}, {Os{\l}owski}, {Reardon}, {van Straten},
  {Toomey}, {Wang}, {Wen}, {Wyithe}, \& {Zhu}}]{Shannon2015}
{Shannon}, R.~M., {Ravi}, V., {Lentati}, L.~T., {et~al.} 2015, Science, 349,
  1522

\bibitem[{{Shapiro}(2013)}]{Shapiro2013}
{Shapiro}, S.~L. 2013, \prd, 87, 103009

\bibitem[{{Shen} {et~al.}(2011){Shen}, {Liu}, {Greene}, \&
  {Strauss}}]{Shen2011}
{Shen}, Y., {Liu}, X., {Greene}, J.~E., \& {Strauss}, M.~A. 2011, \apj, 735, 48

\bibitem[{{Shi} \& {Krolik}(2015)}]{Shi2015}
{Shi}, J.-M., \& {Krolik}, J.~H. 2015, \apj, 807, 131

\bibitem[{{Shi} {et~al.}(2014){Shi}, {Luo}, {Comte}, {Chen}, {Wei}, {Zhao},
  {Wu}, {Zhang}, {Shen}, {Yang}, {Wu}, {Wu}, {Zhang}, {Lei}, {Zhang}, {Wang},
  {Jin}, \& {Zhang}}]{Shi2014}
{Shi}, Z.-X., {Luo}, A.-L., {Comte}, G., {et~al.} 2014, Research in Astronomy
  and Astrophysics, 14, 1234

\bibitem[{{Shields} {et~al.}(2012){Shields}, {Rosario}, {Junkkarinen},
  {Chapman}, {Bonning}, \& {Chiba}}]{Shields2012}
{Shields}, G.~A., {Rosario}, D.~J., {Junkkarinen}, V., {et~al.} 2012, \apj,
  744, 151

\bibitem[{{Silk} \& {Rees}(1998)}]{silk98}
{Silk}, J., \& {Rees}, M.~J. 1998, \aap, 331, L1

\bibitem[{{Simon} \& {Burke-Spolaor}(2016)}]{Simon2016}
{Simon}, J., \& {Burke-Spolaor}, S. 2016, \apj, 826, 11

\bibitem[{{Smith} {et~al.}(2010){Smith}, {Shields}, {Bonning}, {McMullen},
  {Rosario}, \& {Salviander}}]{Smith2010}
{Smith}, K.~L., {Shields}, G.~A., {Bonning}, E.~W., {et~al.} 2010, \apj, 716,
  866

\bibitem[{{Steinborn} {et~al.}(2016){Steinborn}, {Dolag}, {Comerford},
  {Hirschmann}, {Remus}, \& {Teklu}}]{Steinborn2015}
{Steinborn}, L.~K., {Dolag}, K., {Comerford}, J.~M., {et~al.} 2016, \mnras,
  458, 1013

\bibitem[{{Tadhunter} {et~al.}(2012){Tadhunter}, {Almeida}, {Morganti}, {Holt},
  {Rose}, {Dicken}, \& {Inskip}}]{tadhunter12}
{Tadhunter}, C.~N., {Almeida}, C.~R., {Morganti}, R., {et~al.} 2012, \mnras,
  427, 1603

\bibitem[{{Teng} {et~al.}(2012){Teng}, {Schawinski}, {Urry}, {Darg}, {Kaviraj},
  {Oh}, {Bonning}, {Cardamone}, {Keel}, {Lintott}, {Simmons}, \&
  {Treister}}]{teng12}
{Teng}, S.~H., {Schawinski}, K., {Urry}, C.~M., {et~al.} 2012, \apj, 753, 165

\bibitem[{{Thorne} \& {Braginskii}(1976)}]{thorne76}
{Thorne}, K.~S., \& {Braginskii}, V.~B. 1976, \apjl, 204, L1

\bibitem[{{Tokunaga} \& {Vacca}(2005)}]{Tokunaga2005}
{Tokunaga}, A.~T., \& {Vacca}, W.~D. 2005, \pasp, 117, 421

\bibitem[{{Toomre} \& {Toomre}(1972)}]{toomre72}
{Toomre}, A., \& {Toomre}, J. 1972, \apj, 178, 623

\bibitem[{{Tremmel} {et~al.}(2018){Tremmel}, {Governato}, {Volonteri}, {Quinn},
  \& {Pontzen}}]{Tremmel2018}
{Tremmel}, M., {Governato}, F., {Volonteri}, M., {Quinn}, T.~R., \& {Pontzen},
  A. 2018, \mnras, 475, 4967

\bibitem[{{Trias} \& {Sintes}(2008)}]{trias08}
{Trias}, M., \& {Sintes}, A.~M. 2008, \prd, 77, 024030

\bibitem[{{Tsai} {et~al.}(2013){Tsai}, {Jarrett}, {Stern}, {Emonts}, {Barrows},
  {Assef}, {Norris}, {Eisenhardt}, {Lonsdale}, {Blain}, {Benford}, {Wu},
  {Stalder}, {Stubbs}, {High}, {Li}, \& {Kong}}]{Tsai2013}
{Tsai}, C.-W., {Jarrett}, T.~H., {Stern}, D., {et~al.} 2013, \apj, 779, 41

\bibitem[{{Van Wassenhove} {et~al.}(2012){Van Wassenhove}, {Volonteri},
  {Mayer}, {Dotti}, {Bellovary}, \& {Callegari}}]{vanwassenhove12}
{Van Wassenhove}, S., {Volonteri}, M., {Mayer}, L., {et~al.} 2012, \apjl, 748,
  L7

\bibitem[{{Vecchio}(1997)}]{vecchio97}
{Vecchio}, A. 1997, Classical and Quantum Gravity, 14, 1431

\bibitem[{{Veilleux} \& {Osterbrock}(1987)}]{veilleux87}
{Veilleux}, S., \& {Osterbrock}, D.~E. 1987, \apjs, 63, 295

\bibitem[{{Veilleux} {et~al.}(2001){Veilleux}, {Shopbell}, \&
  {Miller}}]{veilleux01}
{Veilleux}, S., {Shopbell}, P.~L., \& {Miller}, S.~T. 2001, \aj, 121, 198

\bibitem[{{Villforth} \& {Hamann}(2015)}]{Villforth2015}
{Villforth}, C., \& {Hamann}, F. 2015, \aj, 149, 92

\bibitem[{{Volonteri} {et~al.}(2003){Volonteri}, {Haardt}, \&
  {Madau}}]{volonteri03}
{Volonteri}, M., {Haardt}, F., \& {Madau}, P. 2003, \apj, 582, 559

\bibitem[{{Wang} {et~al.}(2009){Wang}, {Chen}, {Hu}, {Mao}, {Zhang}, \&
  {Bian}}]{wang09}
{Wang}, J., {Chen}, Y., {Hu}, C., {et~al.} 2009, \apjl, 705, L76

\bibitem[{{Wang} \& {Gao}(2010)}]{wang10}
{Wang}, J., \& {Gao}, Y. 2010, Research in Astronomy and Astrophysics, 10, 309

\bibitem[{{Wang} \& {Yuan}(2012)}]{Wang2012}
{Wang}, X.-W., \& {Yuan}, Y.-F. 2012, \mnras, 427, L1

\bibitem[{{Whelan} \& {Garcia}(2008)}]{Whelan2008}
{Whelan}, E., \& {Garcia}, P. 2008, in Lecture Notes in Physics, Berlin
  Springer Verlag, Vol. 742, Jets from Young Stars II, ed. F.~{Bacciotti},
  L.~{Testi}, \& E.~{Whelan}, 123

\bibitem[{{White} {et~al.}(1997){White}, {Becker}, {Helfand}, \&
  {Gregg}}]{white97}
{White}, R.~L., {Becker}, R.~H., {Helfand}, D.~J., \& {Gregg}, M.~D. 1997,
  \apj, 475, 479

\bibitem[{{Xu} \& {Komossa}(2009)}]{xu09}
{Xu}, D., \& {Komossa}, S. 2009, \apjl, 705, L20

\bibitem[{{York} {et~al.}(2000){York}, {Adelman}, {Anderson}, {Anderson},
  {Annis}, {Bahcall}, {Bakken}, {Barkhouser}, {Bastian}, {Berman}, {Boroski},
  {Bracker}, {Briegel}, {Briggs}, {Brinkmann}, {Brunner}, {Burles}, {Carey},
  {Carr}, {Castander}, {Chen}, {Colestock}, {Connolly}, {Crocker}, {Csabai},
  {Czarapata}, {Davis}, {Doi}, {Dombeck}, {Eisenstein}, {Ellman}, {Elms},
  {Evans}, {Fan}, {Federwitz}, {Fiscelli}, {Friedman}, {Frieman}, {Fukugita},
  {Gillespie}, {Gunn}, {Gurbani}, {de Haas}, {Haldeman}, {Harris}, {Hayes},
  {Heckman}, {Hennessy}, {Hindsley}, {Holm}, {Holmgren}, {Huang}, {Hull},
  {Husby}, {Ichikawa}, {Ichikawa}, {Ivezi{\'c}}, {Kent}, {Kim}, {Kinney},
  {Klaene}, {Kleinman}, {Kleinman}, {Knapp}, {Korienek}, {Kron}, {Kunszt},
  {Lamb}, {Lee}, {Leger}, {Limmongkol}, {Lindenmeyer}, {Long}, {Loomis},
  {Loveday}, {Lucinio}, {Lupton}, {MacKinnon}, {Mannery}, {Mantsch}, {Margon},
  {McGehee}, {McKay}, {Meiksin}, {Merelli}, {Monet}, {Munn}, {Narayanan},
  {Nash}, {Neilsen}, {Neswold}, {Newberg}, {Nichol}, {Nicinski}, {Nonino},
  {Okada}, {Okamura}, {Ostriker}, {Owen}, {Pauls}, {Peoples}, {Peterson},
  {Petravick}, {Pier}, {Pope}, {Pordes}, {Prosapio}, {Rechenmacher}, {Quinn},
  {Richards}, {Richmond}, {Rivetta}, {Rockosi}, {Ruthmansdorfer}, {Sandford},
  {Schlegel}, {Schneider}, {Sekiguchi}, {Sergey}, {Shimasaku}, {Siegmund},
  {Smee}, {Smith}, {Snedden}, {Stone}, {Stoughton}, {Strauss}, {Stubbs},
  {SubbaRao}, {Szalay}, {Szapudi}, {Szokoly}, {Thakar}, {Tremonti}, {Tucker},
  {Uomoto}, {Vanden Berk}, {Vogeley}, {Waddell}, {Wang}, {Watanabe},
  {Weinberg}, {Yanny}, \& {Yasuda}}]{York2000}
{York}, D.~G., {Adelman}, J., {Anderson}, Jr., J.~E., {et~al.} 2000, \aj, 120,
  1579

\bibitem[{{Yu}(2002)}]{yu02}
{Yu}, Q. 2002, \mnras, 331, 935

\bibitem[{{Yu} {et~al.}(2011){Yu}, {Lu}, {Mohayaee}, \& {Colin}}]{yu11}
{Yu}, Q., {Lu}, Y., {Mohayaee}, R., \& {Colin}, J. 2011, \apj, 738, 92

\bibitem[{{Yuan} {et~al.}(2016){Yuan}, {Strauss}, \& {Zakamska}}]{Yuan2016}
{Yuan}, S., {Strauss}, M.~A., \& {Zakamska}, N.~L. 2016, \mnras, 462, 1603

\bibitem[{{Zamanov} {et~al.}(2002){Zamanov}, {Marziani}, {Sulentic}, {Calvani},
  {Dultzin-Hacyan}, \& {Bachev}}]{zamanov02}
{Zamanov}, R., {Marziani}, P., {Sulentic}, J.~W., {et~al.} 2002, \apjl, 576, L9

\bibitem[{{Zhou} {et~al.}(2004){Zhou}, {Wang}, {Zhang}, {Dong}, \&
  {Li}}]{zhou04}
{Zhou}, H., {Wang}, T., {Zhang}, X., {Dong}, X., \& {Li}, C. 2004, \apjl, 604,
  L33

\end{thebibliography}

\end{document}